\documentclass[twocolumn,nofootinbib,longbibliography,prd,superscriptaddress]{revtex4-1}

\usepackage{feynmp}
\usepackage{amsmath}
\usepackage{graphicx}
\usepackage{multirow}
\usepackage{latexsym}
\usepackage{textcomp}
\usepackage{verbatim}
\usepackage{color}
\usepackage{bm}
\usepackage{subfigure}
\usepackage{everysel}
\usepackage{keyval}
\usepackage{ragged2e}
\usepackage{dsfont}
\usepackage{amssymb}
\usepackage{enumitem}
\usepackage{pstricks}
\usepackage{mathtools}
\usepackage{braket}
\usepackage{slashed} 
\usepackage[colorlinks=true]{hyperref} 
\usepackage{tikz}

\usepackage{MnSymbol}

\usepackage{graphicx}
\usepackage{xcolor}
\usepackage{amsmath}
\usepackage{amsfonts}
\usepackage{bbm}

\usepackage{ulem}

\newcommand{\osum}{{%
    \setbox0\hbox{\circ}%
    \rlap{\hbox to \wd0{\hss\sum\hss}}\box0
}}

\def\+{{+\!\!\!+}}		
\def\ppmm{{\overset{+\!\!\!+}{=}} }

\newcommand{\cA}{\mathcal{A} }
\newcommand{\cB}{\mathcal{B} }
\newcommand{\cC}{\mathcal{C} }
\newcommand{\cD}{\mathcal{D} }

\newcommand{\cH}{\mathcal{H} }

\newcommand{\cJ}{\mathcal{J} }

\newcommand{\cL}{\mathcal{L} }
\newcommand{\cM}{\mathcal{M} }
\newcommand{\cN}{\mathcal{N} }
\newcommand{\cO}{\mathcal{O} }
\newcommand{\cP}{\mathcal{P} }
\newcommand{\cQ}{\mathcal{Q} }

\newcommand{\bP}{\mathbb{P} }   

\newcommand{\bW}{\mathbb{W} }

\newcommand{\fpi}{\pi }    

\newcommand{\spi}{\pi }    

\newcommand{\pb}{\text{\tiny PB} }
\newcommand{\dirac}{\text{\tiny D} }

\newcommand{\ttb}{$T\bar{T}$}

\newcommand{\ie}{\textit{i.e.}, }
\newcommand{\eg}{\textit{e.g.}, }

\newcommand{\broken}{\textit{\tiny broken}}

\begin{document}

\title{\ttb \ Deformation of $\cN=(1,1)$ Off-Shell Supersymmetry and Partially Broken Supersymmetry}

\author{Kyung-Sun Lee}
\thanks{Electronic Address: kyungsun.cogito.lee@gmail.com}
\affiliation{School of Physics and Chemistry, Gwangju Institute of Science and Technology, Gwangju 61005, Korea}
\affiliation{School of Physics, Korea Institute for Advanced Study, Seoul 02455, Korea.}

\author{Junggi Yoon}
\thanks{Electronic Address: junggi.yoon@apctp.org}
\affiliation{Asia Pacific Center for Theoretical Physics, POSTECH, Pohang 37673, Korea}
\affiliation{Department of Physics, POSTECH, Pohang 37673, Korea}
\affiliation{School of Physics, Korea Institute for Advanced Study, Seoul 02455, Korea.}

\date{\today}

\begin{abstract} 
We construct the superaction for the \ttb \ deformation of 2D free $\cN=(1,1)$ supersymmetric model with a deformed superfield. We show that the $\cN=(1,1)$ off-shell supersymmetry in the free theory is deformed under the \ttb \ deformation, which is incorporated in the deformed superfield. We interpret this superaction as an effective action of the Goldstone superfield for the partial spontaneous breaking of $\cN=(2,2)$ supersymmetry to $\cN=(1,1)$. We show that the unbroken and broken supersymmetry of the effective superaction corresponds to the off-shell $\cN=(1,1)$ supersymmetry and the off-shell fermi global non-linear symmetry in the \ttb-deformed theory, respectively. We demonstrate that this effective superaction can be obtained by the non-linear realization of the partially broken global supersymmetry~(PBGS) from the coset superspace. Furthermore, we reproduce the superaction by the constrained superfield method accompanied by a field redefinition.
\end{abstract}

\maketitle

\section{Introduction}
\label{sec: introduction}

Quantum field theory~(QFT) is a powerful framework for understanding particles and their interactions. Despite its success in describing a wide range of phenomena, there are still many open questions about the nature of QFT, such as its relation to gravity and its behavior at high energies. One approach to addressing these questions is to study the irrelevant deformation of conformal field theory~(CFT). However, under the renormalization group flow, the irrelevant deformation of CFT, in general, cannot be controlled due to an infinite number of counter terms. Furthermore, irrelevant deformation is often involved with non-local interactions, such as higher-derivative terms. These higher-derivative terms can lead to the emergence of ghosts or tachyons, which are not present in the original theory.

In recent years, there has been significant interest in the study of a special irrelevant deformation of two-dimensional QFT generated by the determinant of the energy-momentum tensor, $\det({T^\mu}_{\nu})$~\cite{Zamolodchikov:2004ce,Smirnov:2016lqw,Cavaglia:2016oda,Jiang:2019epa}. This special irrelevant deformation, so-called \ttb \ deformation, follows a specific flow from IR to UV, along which the special flow, no extra degree of freedom such as ghost does not emerge. Furthermore, \ttb \ deformation features the universal formula for the deformed spectrum, which determines the energy and the momentum of the deformed theory exactly in terms of the undeformed energy and momentum~\cite{Smirnov:2016lqw,Cavaglia:2016oda}.

This universal formula for the deformed spectrum does not distinguish bosonic or fermionic nature of the state, which indicates that the supersymmetry of the undeformed theory will be preserved under the \ttb \ deformation although the \ttb \ deformation operator itself is not supersymmetric. The on-shell supersymmetry has been explicitly shown for the case of the \ttb \ deformation of the free $\cN=(1,1)$ SUSY model~\cite{Lee:2021iut}. However, the \ttb \ deformation of the off-shell supersymmetry transformation has not been fully understood~\cite{Coleman:2019dvf,Lee:2021iut}.

The Refs~\cite{Baggio:2018rpv,Chang:2018dge} proposed the SUSY \ttb \ deformation in the superspace where the superaction is deformed by bilinear of supercurrents. In this SUSY \ttb \ deformation, the off-shell supersymmetry is not deformed since the superfield is not modified under the deformation. It was found in~\cite{Baggio:2018rpv,Chang:2018dge} that the supersymmetric flow equation of the superaction reproduces the original \ttb \ flow equation with the on-shell condition. However in spite of the on-shell equivalence of the flow equation, it was demonstrated in Ref.~\cite{Lee:2021iut} by canonical analysis that the SUSY \ttb-deformed superactions in Refs.~\cite{Baggio:2018rpv,Chang:2018dge} incorporate fermions with higher order time-derivatives. With this higher order time-derivatives the deformed theory does not have the second class constraints anymore in contrast to the undeformed theory with the second class constraints for fermions. Hence, these higher order time-derivative fermions lead to additional degrees of freedom in the phase space of the deformed theory. In Ref.~\cite{Lee:2021iut}, it was shown that these extra degrees of freedom give rise to non-unitarity issue like the Ostrogradsky instability. And this non-unitarity could bring about the potential failure of the factorization of \ttb \ operator in Ref.~\cite{Zamolodchikov:2004ce}. Such an extra degree freedoms could often be removed by a field-redefinition or a Jacobian. However, in a perturbative analysis, we could not find a local field-redefinition or Jacobian which can eliminate the additional degrees of freedom in the SUSY \ttb-deformed model. In this paper, we aim at finding the superaction without higher-derivative fermions which reproduces the \ttb-deformed Lagrangian of the free $\cN=(1,1)$ SUSY model. Surprisingly, we found that a deformed superfield is inevitable to construct such a superaction, which leads to the deformed off-shell supersymmetry. Furthermore, this deformation of the off-shell supersymmetry turns out to be consistent with the non-linear realization of supersymmetry.

Spontaneous symmetry breaking and the effective action for the Goldstone field has been extensively studied in the context of particle physics, condensed matter physics, and cosmology, and has played a pivotal role in the developments of key theoretical frameworks. It was observed that the SUSY \ttb-deformation~\cite{Baggio:2018rpv,Chang:2018dge,Chang:2019kiu,He:2019ahx} of $\cN=(0,2)$ theory~\cite{Jiang:2019hux}, $\cN=(2,2)$ theory~\cite{Ferko:2019oyv} and the complex fermion~\cite{Cribiori:2019xzp} is equivalent to the effective action of the Goldstone superfield. For the case of the \ttb-deformed free $\cN=(1,1)$ SUSY model, it is shown to have the fermi global non-linear symmetry~\cite{Lee:2021iut}. Since this \ttb-deformed theory is related to the $\cN=2$ Green-Schwarz superstring action for 3D target space~\cite{Baggio:2018rpv,Frolov:2019nrr,Lee:2021iut}, this non-linear symmetry was proved to be identical to the broken part of $\cN=2$ super-Poincare symmetry of the 3D target space~\cite{Lee:2021iut}. Therefore, one may also understand the \ttb-deformed $\cN=(1,1)$ SUSY model as an effective theory for the broken supersymmetry.

In this paper, we study the \ttb \ deformation of the free $\cN=(1,1)$ SUSY model to construct the superaction. We demonstrate that the deformed superfield, which is related to the vanilla superfield via field redefinitions, is necessary to establish the superaction. The deformed superfield reflects the \ttb \ deformation of the $\cN=(1,1)$ off-shell supersymmetry, and we obtain the explicit form of the \ttb-deformed $\cN=(1,1)$ off-shell supersymmetry transformation. Furthermore, we confirm that the superaction is identical to the effective superaction of the Goldstone superfield for the partially broken $\cN=(2,2)$ supersymmetry via field redefinition. We clarify that the broken supersymmetry corresponds to the fermi global non-linear symmetry of the \ttb-deformed theory. We also derive the same superaction not only by the non-linear realization method of the broken symmetry but also by the constrained superfield method.

This paper is organized as follows. In Section~\ref{sec: review}, we review the canonical analysis of the \ttb-deformed free $\cN=(1,1)$ SUSY model. In Section~\ref{sec: superaction}, we construct the superaction for the \ttb-deformed theory with deformed superfield, and we study the unbroken and broken off-shell supersymmetry. In Section~\ref{sec: nonlinear}, we discuss the non-linear realization of the broken symmetry to reproduce the superaction. In Section~\ref{sec: constrained superfield}, we derive the superaction again by the constrained superfield method. In Section~\ref{sec: conclusion}, we make concluding remarks.



\section{Review: \texorpdfstring{\ttb}{TTbar} deformation of Free \texorpdfstring{$\cN=(1,1)$}{N=1,1} SUSY Model}
\label{sec: review}

In this section, we review the canonical analysis of the \ttb-deformed free $\cN=(1,1)$ theory in~\cite{Lee:2021iut}. The Lagrangian density of two-dimensional free $\cN=(1,1)$ supersymmetric theory is given by
\begin{align}
    \cL_0\,=\,2\partial_\+\phi\partial_=\phi + S_{\+,=} + S_{=,\+} \ ,\label{eq: undeformed susy lag} 
\end{align}
where $S_{\pm\!\!\pm,\+}$ and $S_{\pm\!\!\pm,=}$ denotes the fermion bilinear
\begin{align}
	 S_{\+\,,\,\mu}\,\equiv\,  i \psi_+ \partial_\mu \psi_+\ ,\qquad
	S_{=\,,\,\mu}\,\equiv\,  i \psi_- \partial_\mu \psi_-\ . \label{eq: fermion bilinears}
\end{align}
The conventions that we use are summarized in Appendix~\ref{app: convention}. The \ttb \ deformation is characterized by the special flow equation~\cite{Smirnov:2016lqw} given by
\begin{align}
    \partial_\lambda \cL \,=\, {1\over 2} \epsilon_{\mu\nu }\epsilon^{\rho\sigma}{T^\mu}_\rho {T^\nu}_\sigma\ , \label{eq: flow equation}
\end{align}
where the initial condition for the flow equation is the undeformed one~\eqref{eq: undeformed susy lag}, $\cL|_{\lambda=0}=\cL_0$. It is well-known that the improvement term of the energy-momentum tensor does not have any effect on the physical consequences. However, the solution of the flow equation does depend on the improvement term of the energy-momentum tensor on the right-hand side of Eq.~\eqref{eq: flow equation}. Namely, depending on the improvement term, the deformed Lagrangian as a solution of the flow equation could have higher derivatives, which leads to the larger Hilbert space than that of the undeformed one.

For the case of the free $\cN=(1,1)$ model, we use Noether energy-momentum tensor without any improvement term to prevent the inflow of additional degrees of freedom by the higher derivatives~\cite{Lee:2021iut}. And the deformed Lagrangian as a solution of the flow equation~\eqref{eq: flow equation} is found\footnote{The same deformed Lagrangian density of $\cN=(1,1)$ theory can also be obtained by the dynamical coordinate transformation~\cite{Coleman:2019dvf}, Green-Schwarz~(GS)-like action with uniform light-cone gauge~\cite{Baggio:2018rpv,Frolov:2019nrr,Frolov:2019xzi} or GS-like action with static gauge \cite{Lee:2021iut}.} to be
\begin{equation}
\begin{aligned}
	\mathcal{L}&\,=\,-{1\over 2\lambda} \left[ \sqrt{1 + 2\chi } -1 \right]\cr
    &\quad+ { 1 +\chi + \sqrt{1  +2 \chi } \over 2 \sqrt{1+ 2\chi  } } (S_{\+,=}+S_{=,\+}) \cr
	&\quad+{2\lambda \over \sqrt{1+ 2 \chi }} [ (\partial_=\phi)^2 S_{\+,\+}+ (\partial_\+ \phi)^2 S_{=,=} ] \cr
	&\quad + \lambda{ 1+ \chi -\chi^2 + ( 1+ 2\chi )^{3\over 2}  \over 2 ( 1+ 2\chi )^{3\over 2}}S_{\+,\+}S_{=,=}\cr
    &\quad - \lambda{ 1 + 3 \chi + \chi^2 + ( 1+ 2\chi )^{3\over 2}  \over 2 ( 1+ 2\chi )^{3\over 2}}S_{\+,=}S_{=,\+} \cr
	&\quad- {2\lambda^2 \chi \over  ( 1+ 2\chi )^{3\over 2} }\cr
    &\quad\times[ (\partial_\+\phi)^2 S_{\+,=} S_{=,=} + (\partial_= \phi)^2 S_{\+,\+}S_{=,\+} ]\ ,\quad \label{eq: susy lag}
\end{aligned}
\end{equation}
where $\chi \, \equiv\,   -  4  \lambda \partial_\+\phi \partial_=\phi$. Moreover, the Hamiltonian density and momentum density of the deformed theory can be written as 
\begin{align}
    \begin{split}
    \cH\, = \, & {1\over 2\lambda }\left[\sqrt{1+4\lambda \cH_{b,0} +4\lambda^2 \cP_{b,0}^2}-1 \right]\cr
    & + {1\over 2} \left({1-4\lambda^2 \cP_{b,0}^2\over \sqrt{1+ 4\lambda \cH_{b,0} +4\lambda^2 \cP_{b,0}^2}}+1 \right)\cH_{f,0} \cr
    & + \lambda \cP_{b,0}\cP_{f,0}\cr
    &- {2\lambda^3(\cH_{b,0}^2-\cP_{b,0}^2) \over (1+ 4\lambda \cH_{b,0} + 4\lambda^2 \cP_{b,0}^2)^{3\over 2} }(\cH_{f,0}^2-\cP_{f,0}^2)\ ,\label{eq: ham op susy}
    \end{split}\\
    \begin{split}
    \cP\,=\, & \left(1-\lambda \cH_{f,0}\right) \cP_{b,0}\cr
    &+{1\over 2}\left[1+\sqrt{1+ 4\lambda \cH_{b,0} +4\lambda^2 \cP_{b,0}^2}\right]\cP_{f,0} \ ,\label{eq: momentum op susy}
    \end{split}
\end{align}
where $\cH_{b/f,0}$ and $\cP_{b/f,0}$ denotes the Hamiltonian density and momentum density of free bosonic field and fermionic field, respectively:
\begin{align}
    &\cH_{b,0}\equiv  {1\over 2} \spi^2 +{1\over 2} \phi'^2\ ,  &&\cP_{b,0} \equiv  \spi \phi'\ , \\
    &\cH_{f,0} \equiv {i\over2} \psi_+\psi'_+-{i\over2} \psi_-\psi'_-\ , &&\cP_{f,0} \equiv {i\over2} \psi_+\psi'_++{i\over2} \psi_-\psi'_-\ .
\end{align}
The deformed Lagrangian~\eqref{eq: susy lag} is linear in the time derivative of the fermion $\dot{\psi}_\pm$. Therefore, the variation with respect to $\dot{\psi}_\pm$ gives us the constraints like free fermion case. If the Lagrangian had higher order in $\dot{\psi}_\pm$, we would not have had the constraints, and in turn, we would have encountered the larger Hilbert space. It was shown~\cite{Lee:2021iut} that the deformed (second class) constraint can be written in terms of deformed Hamiltonian density $\cH$ \eqref{eq: ham op susy} and deformed momentum density $\cP$ \eqref{eq: momentum op susy} as
\begin{align}
    \cC_1\,=\,& \fpi_+ -{i\over 2}\psi_+ - {i\over 2}\lambda (\cH -\cP) \psi_+\ , \label{eq: susy const p1}\\
    \cC_2\,=\,& \fpi_- -{i\over 2}\psi_- - {i\over 2}\lambda (\cH +\cP) \psi_-\ .\label{eq: susy const p2}
\end{align}

From the second-class constraints \eqref{eq: susy const p1} and \eqref{eq: susy const p2}, we can evaluate the Dirac bracket defined as

\begin{align}
	 &\{F(x),G(y)\}_\dirac\cr
     &\,\equiv\,\{F(x),G(y)\}_\pb-  \sum_{i,j=1,2}\bigg(\int dzdw\; \{F(x),\cC_i(z)\}_\pb\cr
	 &\qquad \times\cM^{-1}(i,z;j,w)\{\cC_j(w),G(y)\}_\pb\bigg)\ , \label{eq: dirac bracket}
\end{align}

with the matrix $\cM(i,x;j,y)$  
\begin{align}
    \cM(i,x;j,y)\,\equiv\, \{\cC_i(x),\cC_j(y)\}_\pb\ .
\end{align}
Here, the Poisson bracket of the system involving the scalar and fermion fields $\phi,\psi_\pm$ and its conjugate momenta $\spi, \fpi_\pm$ is defined by
\begin{equation}
\begin{aligned}
    &\{F(x),G(y)\}_\pb\cr
    &\equiv  \int dz\;\bigg[ \bigg({\partial F(x) \over \partial \phi(z) }{\partial G(y) \over \partial \spi(z) }- {\partial F(x) \over \partial \spi(z) }{\partial G(y) \over \partial \phi(z) }\bigg)  \cr
	&+\sum_{\alpha=\pm }\bigg({F(x) \overleftarrow{\partial}\over \overleftarrow{\partial} \psi_\alpha(z) }{\overrightarrow{\partial}G(y) \over \overrightarrow{\partial} \fpi_\alpha(z) }+ {F(x) \overleftarrow{\partial}\over \overleftarrow{\partial} \fpi_\alpha(z) }{\overrightarrow{\partial}G(y) \over \overrightarrow{\partial} \psi_\alpha(z) }\bigg)\bigg]\ .\label{def: susy poisson bracket}
\end{aligned}
\end{equation}

The \ttb-deformed free $\cN=(1,1)$ model is shown to have global symmetry~\cite{Lee:2021iut}. And the charges of this global symmetry are evaluated by the Noether procedure:
\begin{align}
    Q^1_\pm\, \equiv \, &\int dx\; \psi_\pm(\spi\pm\phi')\ , \label{eq: supercharge ttb susy}\\
    Q^2_\pm\,\equiv \,&- {8\pi i\over L} \int dx\; \fpi_\pm\ ,\label{eq: broken supercharge ttb susy}\\
    \bP^2\, \equiv&\, {2\pi \over L}\int dx \; \spi\ . \label{eq: scalar shift ttb}
\end{align}
where $L$ denotes the circumference of the compact spatial coordinate $x$. Note that the $\lambda$-dependencies are inherited from the conjugate momenta $\spi, \fpi_\pm$ in Eqs.~\eqref{eq: susy const p1} and \eqref{eq: susy const p2}. The fermionic charge $Q^1_\pm$ in Eq.~\eqref{eq: supercharge ttb susy} is, in fact, the supercharges of the deformed $\cN=(1,1)$ supersymmetry given by

\begin{align}
    \begin{split}
    \delta^1_\pm \phi\,=\,&-{1\over 2} \left(1+\sqrt{1+2\lambda \phi'^2\over 1+2\lambda \spi^2}\right)\psi_\pm\pm {i\lambda^2 \spi (\spi\pm\phi')\over 2(1+2\lambda\spi^2) }\cr
    &\times\left( 1+{1\pm2\lambda \spi \phi'\over \sqrt{(1+2\lambda \spi^2)(1+2\lambda \phi'^2)}} \right) \psi_\mp\psi'_\mp\psi_\pm\ ,\label{eq: canonical phi susy var}
    \end{split}
    \\
    \begin{split}
    \delta^1_\pm \psi_\pm\,=\,&{-1\pm2\lambda \spi \phi' +\sqrt{(1+2\lambda \spi^2)(1+2\lambda \phi'^2)} \over i\lambda (\spi \pm\phi')}\cr
    &+ \lambda \bigg(\pm {\phi'^2 +\spi^2 +4 \lambda \spi^2 \phi'^2 \over (\spi \pm \phi') \sqrt{(1+2\lambda \spi^2)(1+2\lambda \phi'^2) } }\cr
    &+{2\spi \phi' \over \spi \pm\phi'}\bigg)\psi_\mp\psi'_\mp\cr
    &+{i\lambda^2 (\spi \pm \phi')\over 2 (1+2\lambda \spi^2)(1+2\lambda \phi'^2)}\psi_+\psi'_+\psi_- \psi'_-\cr
    &\times (1\pm 2\lambda \spi \phi' + \sqrt{(1+2\lambda \spi^2)(1+2\lambda \phi'^2)})\ ,\label{eq: canonical psi susy var1}
    \end{split}
    \\
    \begin{split}
    \delta^1_\pm \psi_\mp \,=\,&\mp{\lambda(\spi \pm \phi') \over \sqrt{(1+2\lambda \spi^2)(1+2\lambda \phi'^2)}}\psi_\pm\psi'_\mp\ , \label{eq: canonical psi susy var2}
    \end{split}
\end{align}
where we used $\delta^{1,2}_\pm (\cdot)\,\equiv\,i\{Q^{1,2}_\pm,\cdot\}_\dirac$.

In undeformed theory ($\lambda=0$), the charge $Q^2_\pm$ and $\bP^2$ in eq.~\eqref{eq: broken supercharge ttb susy} and \eqref{eq: scalar shift ttb} generates the shift of fermi field and scalar field, respectively. However since the Dirac bracket~\eqref{eq: dirac bracket} is modified by \ttb \  deformation, the shifting symmetry generated by $Q^2_\pm$ is also deformed:
\begin{align}
    \begin{split}
    \delta^2_\pm \phi &={4\pi \over L}\bigg[\pm{i\lambda \over 2}\left(\phi'\mp\spi\sqrt{{1+2\lambda\phi'^2 \over 1+2\lambda\spi^2  }}\right)\\ 
    &\quad\pm{\lambda^2(\spi\pm\phi')\over 4(1+2\lambda\spi^2)\sqrt{(1+2\lambda\spi^2)(1+2\lambda\phi'^2)}}\\
    &\quad\times (1\pm2\lambda\spi\phi'+\sqrt{(1+2\lambda\spi^2)(1+2\lambda\phi'^2)})\\
    &\quad\times\psi_\mp\psi'_\mp\bigg]\psi_\pm\ , \label{eq: canonical phi broken susy var}
    \end{split}\\
    \begin{split}
    \delta^2_\pm\psi_\pm&={4\pi \over L}\bigg[1\mp{i\lambda \psi_\pm\psi'_\pm\over 2\sqrt{(1+2\lambda\spi^2)(1+2\lambda\phi'^2)}}\\ 
    &\quad\times (1\pm2\lambda\spi\phi'-\sqrt{(1+2\lambda\spi^2)(1+2\lambda\phi'^2)})\\
    &\quad+{\lambda^2(1+\lambda(\spi^2+\phi'^2))\psi_+\psi'_+\psi_-\psi'_-\over 2(1+2\lambda\spi^2)^{3/2}(1+2\lambda\phi'^2)^{3/2}}\\
    &\quad\times (1\pm2\lambda\spi\phi'+\sqrt{(1+2\lambda\spi^2)(1+2\lambda\phi'^2)})\bigg]\ , \label{eq: canonical psi broken susy var1}
    \end{split}\\
    \begin{split}
    \delta^2_\pm\psi_\mp&=\pm{4\pi \over L}\bigg[{i\lambda \psi_\pm\psi'_\mp\over2\sqrt{(1+2\lambda\spi^2)(1+2\lambda\phi'^2)}}\\
    &\quad\times (1\mp2\lambda\spi\phi'+\sqrt{(1+2\lambda\spi^2)(1+2\lambda\phi'^2)})\bigg]\ . \label{eq: canonical psi broken susy var2}
    \end{split}
\end{align}
On the other hand, the deformed symmetry generated by $\bP^2$ remains to be the shift symmetry of scalar field $\phi$:
\begin{equation}
    \{\bP^2,\phi\}_\dirac \,=\,  -1\ ,\quad \{\bP^2,\psi_\pm\}_\dirac \,=\,  0\ .
\end{equation}
The algebra of the fermi charges with respect to the deformed Dirac bracket was evaluated in~\cite{Lee:2021iut} to be
\begin{align}
    i\{Q^1_\pm,Q^1_\pm\}_\dirac&\,=\, 2(H\pm P)\\
    i\{Q^2_\pm,Q^2_\pm\}_\dirac&\, =\,  {16\pi^2\over L} + {16\pi^2 \lambda\over L^2} (H\mp P)\ ,\label{eq: sq algebra4}\\
    i\{Q^1_\pm,Q^2_\pm\}_\dirac&\, =\, 2\left(\bP^2 \pm {4\pi^2\over L^2}\bW^2\right)\ ,  \\
    \{Q^1_\pm,Q^1_\mp\}_\dirac&\,= \,\{Q^2_\pm,Q^2_\mp\}_\dirac\,= \, \{Q^1_\pm,Q^2_\mp\}_\dirac\, =\, 0\ ,
\end{align}
where the Hamiltonian $H$, momentum $P$ and the topological charge $\bW^2$ which is the winding number of the scalar field $\phi$ are defined by
\begin{align}
    H=   \int dx \cH\ ,\quad P =   \int dx \cP\ ,\quad \bW^2\equiv {L\over 2\pi }\oint dx  \phi' \ .
\end{align}
The global charges $Q^{1,2}_\pm, \bP^2$ commute with the Hamiltonian and momentum with respect to the deformed Dirac bracket:
\begin{equation}
\begin{aligned}
    &\{Q^{1,2}_\pm,H\}_\dirac\,=\,\{Q^{1,2}_\pm,P\}_\dirac\,=\,0\ ,\\
    &\{\bP^2,H\}_\dirac\,=\,\{\bP^2,P\}_\dirac\,=\,0\ .\label{eq: dirac brackets of h p}
\end{aligned}
\end{equation}
It was demonstrated~\cite{Lee:2021iut} that they correspond to the charges of the $\cN=2$ super Poincare symmetry of 3D target space in 3D $\cN=2$ GS-like model \cite{Mezincescu:2011nh}. The discrete lightcone quantization~(DLCQ) was used in relating the 3D $\cN=2$ GS-like model to the \ttb-deformed $\cN=(1,1)$ SUSY model~\cite{Lee:2021iut}. And the topological charge due to the DLCQ breaks the $\cN=2$ super Poincare symmetry of the 3D target space which is identical to the algebra of the charges in the \ttb-deformed one. The broken supersymmetry corresponds to the fermi global charge $Q^2_\pm$ while the unbroken ones are identified to be the deformed $\cN=(1,1)$ supercharge $Q^1_\pm$ of the \ttb-deformed theory~\cite{Lee:2021iut}.

In this paper, we present another interpretation on this algebra as partially broken global supersymmetry~(PBGS) of 2D SQFT where $Q^1_\pm$ are the unbroken supersymmetry while $Q^2_\pm$ are broken one. Accordingly, the \ttb-deformed Lagrangian~\eqref{eq: susy lag} can be understood as an effective Lagrangian with non-linearly realized supersymmetry.

\section{Superaction for \texorpdfstring{\ttb}{TTbar}-deformed Lagrangian}
\label{sec: superaction}

The deformed energy $E(\lambda)$ and the deformed momentum $P(\lambda)$ of the \ttb-deformed theory defined on the cylinder of radius $L$ is universally expressed in terms of the undeformed energy $E_{(0)}$ and the undeformed momentum $P_{(0)}$~\cite{Smirnov:2016lqw,Cavaglia:2016oda}:
\begin{align}
    E(\lambda)\,=\, &{L\over 2\lambda }\bigg[ \sqrt{1+ {4\lambda \over L}E_{(0)}+ {4\lambda^2\over L^2}P^2_{(0)}}-1\bigg]\ ,\\
    P(\lambda)\,=\,&P_{(0)} \ .
\end{align}
This universal formula for the deformed energy and momentum does not distinguish whether the state is bosonic or fermionic. This implies that the Bose-Fermi degeneracy of the undeformed supersymmetric model will be preserved along the \ttb \ deformation. Therefore, one can expect the supersymmetry of the undeformed theory would be preserved under the \ttb \ deformation, but this looks non-trivial because the \ttb \ flow equation~\eqref{eq: flow equation} is not manifestly supersymmetric.

It was explicitly shown in~\cite{Lee:2021iut} that the supersymmetry of the free $\cN=(1,1)$ SUSY model is preserved along the \ttb \ deformation because the Hamiltonian and the momentum operator is expressed as
\begin{align}
    H\,=\, & {i\over 4} \{Q^1_+,Q^1_+\}_\dirac +{i\over 4} \{Q^1_-,Q^1_-\}_\dirac\ ,\\
    P\,=\, & {i\over 4} \{Q^1_+,Q^1_+\}_\dirac -{i\over 4} \{Q^1_-,Q^1_-\}_\dirac\ .
\end{align}
Thus, it is natural to ask whether we can rewrite the deformed Lagrangian~\eqref{eq: susy lag} in the superspace. In this section, we start with the ansatz for the superaction, and we find the explicit form of superaction which will reproduce the deformed Lagrangian~\eqref{eq: susy lag} after we integrate out the auxiliary field.

To make an ansatz for the superaction, we first need to analyze the mass dimensions of the ingredients\footnote{Here we assume that the only dimensionful parameter in the superaction is the deformation parameter~$\lambda$.}. The mass dimension of superaction $\cA$, deformation parameter $\lambda$, $\cN=(1,1)$ scalar superfield $\Phi$, spacetime derivatives $\partial_{\pm\pm}$ and super-covariant derivatives $D_{\pm}$ are given by
\begin{align}
    [\lambda]=-2\ ,\  [\Phi]=0\ ,\  [\mathcal A]=[\partial_{\pm\pm}]\,=\,1\ ,\  [\cD_{\pm}]=\frac12\ .
\end{align}
In addition to $\cD_-\Phi \cD_+\Phi$, there are 4 Lorentz invariant quadratic terms
\begin{equation}
    \begin{aligned}
    & A\,\equiv\,\partial_{\+}\Phi\partial_{=}\Phi\, ,\quad B\,\equiv\, (\cD_-\cD_+\Phi)^2\, ,\\
    & C\,\equiv\, \cD_+\Phi\partial_{=}\cD_+\Phi\, ,\quad D\,\equiv\, \cD_-\Phi\partial_{\+}\cD_-\Phi\ ,
    \end{aligned}
\end{equation}

and their mass dimension is given by
\begin{equation}
    [A]\,=\,[B]\,=\,[C]\,=\,[D]\,=\,2\ .
\end{equation}
%
%
%
Note that the only possible Lorentz invariant quadratic term with odd mass dimension is $\cD_-\Phi \cD_+\Phi$. Therefore, the ansatz for the superaction of mass dimension 1 should be of form
\begin{equation}
	\mathcal A\,=\,\Omega(\lambda A,\lambda B,\lambda C,\lambda D)\, \cD_-\Phi \cD_+\Phi\ ,
\end{equation}
where $\Omega$ contains only dimensionless quadratics. Furthermore, Because the ansatz is proportional to $\cD_-\Phi \cD_+\Phi$, any dependence of $C= \cD_+\Phi\partial_{=}\cD_+\Phi$ and $D=\cD_-\Phi\partial_{\+}\cD_-\Phi$ in $\Omega$ will vanish. Thus the superaction ansatz can effectively be written as
\begin{equation}
	\mathcal A\,=\,\Omega(\lambda A,\lambda B)\, \cD_-\Phi \cD_+\Phi\ . \label{eq: superaction ansatz}
\end{equation}

\subsection{Failure of constructing superaction by using vanilla superfield}
\label{sec: normal superfield superaction}

Now we attempt to reproduce the deformed Lagrangian \eqref{eq: susy lag} perturbatively from the superaction ansatz~\eqref{eq: superaction ansatz} with the vanilla $\cN=(1,1)$ scalar superfield:
\begin{equation}
    \Phi\,=\,\phi + i\theta^+\psi_+ + i\theta^-\psi_- + i\theta^+\theta^- F\ . \label{eq: usual superfield}
\end{equation}
For this, we perturbatively expand $\Omega(\lambda A, \lambda B)$ in the ansatz~\eqref{eq: superaction ansatz} with respect to $\lambda$
\begin{equation}
\begin{aligned}
    &\Omega(\lambda A,\lambda B)\\
    &\,=\,\Omega^{(0)}+\lambda\,\Omega^{(1)}(A,B)+\lambda^2\,\Omega^{(2)}(A,B)+\cO(\lambda^3)\ ,
\end{aligned}
\end{equation}
where $\Omega^{(0)}$ is a constant number and $\Omega^{(n)}(A,B)$ can be written as $n$-th order homogeneous polynomial of $A$ and $B$ as
\begin{align}
    \Omega^{(1)}(A,B)\,=\, &\Omega_A A+\Omega_B B\ ,\\
    \Omega^{(2)}(A,B)\,=\, &\Omega_{A^2} A^2+\Omega_{ AB} AB+\Omega_{B^2} B^2\quad \mbox{\textit{etc.}}
\end{align}
We will determine $\Omega^{(i)}$s order by order by comparing with the perturbative expansion of the deformed Lagrangian~\eqref{eq: susy lag}, $\cL=\cL^{(0)}+\lambda\cL^{(1)}+\lambda^2\cL^{(2)}+\dots$. For example, $\cL^{(0)}$ and $\cL^{(1)}$ are given by
\begin{align}
\cL^{(0)}&\,=\,2\partial_{\+}\phi\partial_{=}\phi+S_{\+,=}+S_{=,\+}\ , \label{eq: 0th order deformed lagrangian}\\
\begin{split}
\cL^{(1)}&\,=\,4(\partial_{\+}\phi\partial_{=}\phi)^2+2(\partial_{\+}\phi)^2S_{=,=}+2(\partial_{=}\phi)^2S_{\+,\+}\cr
&\quad+ S_{\+,\+}S_{=,=} - S_{\+,=}S_{=,\+} \ . \label{eq: 1st order deformed lagrangian}
\end{split}
\end{align}

\noindent \textbf{$\lambda^0$-order:} At $\lambda^0$ order, the Lagrangian from the superaction ansatz is 
\begin{equation}
\begin{aligned}
    &\int d^2\theta\; \cA^{(0)}\\
    &\,=\,  2\Omega^{(0)} \left(2\partial_{\+}\phi\partial_{=}\phi+i\psi_+\partial_{=}\psi_++i\psi_-\partial_{\+}\psi_-+\frac12 F^2\right)\; , \label{eq: 0th order superaction ansatz}
\end{aligned}
\end{equation}
and the equations of motion for the auxiliary field $F=F^{(0)}+ \cO(\lambda)$ at order $\cO(\lambda^{0})$ is given by
\begin{equation}
    F^{(0)}\,=\,0\ .
\end{equation}
Inserting $F=0+\cO(\lambda)$ to Eq.~\eqref{eq: 0th order superaction ansatz}, one can determine $\Omega^{(0)}={1\over 2}$ to identify with the undeformed Lagrangian $\cL^{(0)}$ in Eq.~\eqref{eq: 0th order deformed lagrangian}.

\vspace{3mm}

\noindent \textbf{$\lambda^1$-order:} At order $\cO(\lambda)$, we have
\begin{align}
    &\int d^2\theta\ \left(\cA^{(0)}+\lambda \cA^{(1)}\right)\cr
    &\,=\, \int d^2\theta\ \left(\Omega^{(0)}+\lambda\,\Omega^{(1)}(A,B)\right)\,  \cD_-\Phi \cD_+\Phi\cr
    &\,=\,  \left(\cL^{(0)}+{1\over2}F^2\right)\cr 
    &\quad+\lambda \Omega_A\bigg(4(\partial_\+\phi\partial_=\phi)^2+2(\partial_\+\phi)^2S_{=,=}+2(\partial_=\phi)^2S_{\+,\+}\cr
    &\qquad\qquad\quad+4\partial_\+\phi\partial_=\phi (S_{\+,=}+S_{=,\+})\qquad\cr
    &\qquad\qquad\quad+S_{\+,\+}S_{=,=} + S_{\+,=}S_{=,\+} + \partial_\+\phi\partial_=\phi\, F^2\bigg)\cr
    &\quad-\lambda \Omega_B\bigg(8S_{\+,=}S_{=,\+}+F^4\cr
    &\qquad\qquad\quad+\left(4\partial_\+\phi\partial_=\phi+6S_{\+,=}+6S_{=,\+}\right)F^2\bigg) \ . \label{eq: 1st order ansatz expansion}
\end{align}
Note that there is no time-derivative acting on $F$. Hence we can again obtain the classical equations of motion for $F$ up to order $\cO(\lambda)$. However, since the classical solution for $F$ is of order $\cO(\lambda)$, it does not give any contribution to the Lagrangian up to order $\cO(\lambda)$ when we insert the solution into Eq.~\eqref{eq: 1st order ansatz expansion}. Therefore, after integrating out the auxiliary field $F$, we have
\begin{align}
     &\left.\left(\int d^2\theta\ \cA\right)\right\vert_{\text{integrate out $F$}}\cr 
     &\,=\,\cL^{(0)}\cr
     &\quad+\lambda \Omega_A\bigg(4(\partial_\+\phi\partial_=\phi)^2+2(\partial_\+\phi)^2S_{=,=}+2(\partial_=\phi)^2S_{\+,\+}\cr
     &\qquad\qquad\quad+4\partial_\+\phi\partial_=\phi (S_{\+,=}+S_{=,\+})\cr
     &\qquad\qquad\quad+S_{\+,\+}S_{=,=} + S_{\+,=}S_{=,\+}\bigg)\cr
     &\quad-\lambda \Omega_B\bigg(8S_{\+,=}S_{=,\+}\bigg) +\cO(\lambda^2)\ .  \label{eq: upto 1st order component form with normal superfield}
\end{align}
Due to the term $4\partial_\+\phi\partial_=\phi (S_{\+,=}+S_{=,\+})$ at order $\cO(\lambda)$, we find that the Lagrangian~\eqref{eq: upto 1st order component form with normal superfield} at order $\cO(\lambda)$ from the superaction ansatz cannot be matched with $\cL^{(1)}$ in Eq.~\eqref{eq: 1st order deformed lagrangian} for any value of $\Omega_A,\,\Omega_B$.

Therefore, we conclude that the \ttb-deformed Lagrangian~\eqref{eq: susy lag} cannot be constructed from the superaction with the vanilla superfield~\eqref{eq: usual superfield}.

\subsection{Construction of superaction with deformed superfield}
\label{sec: Construction of superaction with deformed superfield}

In Section~\ref{sec: review}, we have seen that the deformed Dirac bracket leads to the \ttb-deformed on-shell supersymmetry transformations~\eqref{eq: canonical phi susy var}$\sim$\eqref{eq: canonical psi susy var2} in spite of the same form of the supercharge \eqref{eq: supercharge ttb susy} as that of the undeformed one in terms of canonical variables. This hints at the possibility that the off-shell supersymmetry transformation could also be deformed. In this case, one should use the modified superfield which incorporates the deformed supersymmetry transformation. Therefore we take the following ansatz for the deformed superfield to construct the superaction.
\begin{equation}
\begin{aligned}
	\Phi&\,=\,\phi+i\theta^+(\mathbf f[\phi,\psi_\pm;\lambda]\,\psi_+)+i\theta^-(\mathbf g[\phi,\psi_\pm;\lambda]\,\psi_-)\cr
 &\quad+i\theta^+\theta^-F\ , \label{eq: modified superfield ansatz}
\end{aligned}
\end{equation}
where $\mathbf f, \mathbf g$ are Grassmann even and dimensionless. With this deformed superfield ansatz, we will repeat the same perturbative procedure in the previous section to find the superaction~\eqref{eq: superaction ansatz} which reproduces the \ttb-deformed Lagrangian~\eqref{eq: susy lag}. This time, we also have to determine $\mathbf{f}[\phi,\psi_\pm;\lambda]$ and $\mathbf{g}[\phi,\psi_\pm;\lambda]$. We assume that $\mathbf{f}$ and $\mathbf{g}$ do not include the higher-order derivatives. Then, in the light of the Lorentz covariance and the mass dimension, one can expand $\mathbf{f}$ and $\mathbf{g}$ with respect to the fermion bilinears as follow.
\begin{align}
    \mathbf f\,=\, &\mathbf{f_1}(\chi) + \lambda \mathbf{f_2}(\chi)S_{=,\+} + \lambda^2 \mathbf{f_3}(\chi)(\partial_\+\phi)^2S_{=,=}\ , \\
    \mathbf g\,=\, &\mathbf{g_1}(\chi) + \lambda \mathbf{g_2}(\chi)S_{\+,=} + \lambda^2 \mathbf{g_3}(\chi)(\partial_=\phi)^2S_{\+,\+}\ ,  
\end{align}
where $\chi \equiv -  4  \lambda \partial_\+\phi \partial_=\phi$. Note that other fermion bilinears do not appear because $\mathbf f$ and $\mathbf g$ in \eqref{eq: modified superfield ansatz} is multiplied to $\psi_+$ and $\psi_-$, respectively.

In the same way, as in Section~\ref{sec: normal superfield superaction} with the deformed superfield ansatz~\eqref{eq: modified superfield ansatz}, we determine the superaction ansatz perturbatively by comparing with the \ttb-deformed Lagrangian~\eqref{eq: susy lag}. Up to order $\cO(\lambda^6)$, we could fix the terms in $\Omega(\lambda A,\lambda B)$ that are independent of $B$. 
\begin{equation}
\begin{aligned}
    &\Omega(\lambda A,\lambda B)\\
    &\,=\,\bigg( {1\over2}+\lambda A+4(\lambda A)^2+20(\lambda A)^3+112(\lambda A)^4\\
    &\quad+672(\lambda A)^5+4224(\lambda A)^6 +\cO\left((\lambda A)^7\right)\bigg)\\
    &\quad+(\text{terms with $B$})\ , \label{eq: ansatz factor series}
\end{aligned}
\end{equation}
Furthermore, we could only determine the terms in $\mathbf f, \mathbf g$ which are independent of the fermions.
\begin{equation}
\begin{aligned}
    \mathbf f \,=\, \mathbf g&\,=\, 1+{\chi\over2}-{\chi^2\over4}+{\chi^3\over4}-{5\chi^4\over16}+{7\chi^5\over 16}-{21\chi^6\over32}\cr
    &\quad+\cO(\chi^7) + (\text{terms with fermions})\ , \label{eq: f, g factor series} 
\end{aligned}
\end{equation}
The other terms in $\Omega$ and $\mathbf{f},\mathbf{g}$ that are involved with $B$ or fermions are not uniquely fixed\footnote{We could obtain the equations for the coefficients of those terms, but there is no unique solution for them.} by matching with the deformed Lagrangian~\eqref{eq: susy lag}.

Based on the perturbative solution for the ansatz~\eqref{eq: ansatz factor series}, \eqref{eq: f, g factor series}, we conjecture that the term in $\Omega$ which is independent of $B$ is given by
\begin{equation}
    \Omega(\lambda A,\lambda B)\,=\,{ 1 \over 1+\sqrt{1-8\lambda A } }+(\text{terms with $B$})\ , \label{eq: ansatz factor sqrt}
\end{equation}
and the term in $\mathbf{f}$ and $\mathbf{g}$ which is independent of fermions is 
\begin{align}
    \mathbf f \,=\, & {\sqrt{1+2\chi}+1 \over 2} + (\text{terms with fermions})\ , \label{eq: f factor sqrt}\\
    \mathbf g\,=\, & {\sqrt{1+2\chi}+1 \over 2} + (\text{terms with fermions})\ . \label{eq: g factor sqrt}
\end{align}

To determine the rest of part in Eq.~\eqref{eq: ansatz factor sqrt}, we assume that they are incorporated in $\Omega$ in a minimal way that the dependence on $A$ and $B$ comes only from the term
\begin{equation}
    \cD_-\cD_+(\cD_-\Phi \cD_+\Phi)\,=\,4A-B+2i(C+D)\ . \label{eq: form in sqrt}
\end{equation}
Then, $\Omega$ can be written as
\begin{equation}
\begin{aligned}
    \Omega(\lambda A,\lambda B)\,=\,&{ 1 \over 1+\sqrt{1-2\lambda(4A-B+2i(C+D)) } }\\
    \,=\,&{ 1 \over 1+\sqrt{1-2\lambda \cD_-\cD_+(\cD_-\Phi \cD_+\Phi) }}\ , \label{eq: superaction factor form}
\end{aligned}
\end{equation}
and accordingly the superaction~\eqref{eq: superaction ansatz} becomes
\begin{equation}
    \cA\,=\,{ \cD_-\Phi \cD_+\Phi \over 1+\sqrt{1-2\lambda \cD_-\cD_+(\cD_-\Phi \cD_+\Phi) }}\ . \label{eq: superaction form}
\end{equation}
Once again note that the dependence on $C$ and $D$ in $\Omega$~\eqref{eq: superaction factor form} will vanish in the superaction because of the factor $\cD_-\Phi \cD_+\Phi$ in the superaction~\eqref{eq: superaction form}. In addition, we determine $\mathbf{f}$ and $\mathbf{g}$ to be
\begin{align}
    \begin{split}
	\mathbf f&\,=\,\frac{\sqrt{1+2\chi}+1}{2}-\lambda\left(\frac{1+\chi+\sqrt{1+2\chi}}{2\sqrt{1+2\chi}}\right) S_{=,\+}\cr
    &\quad-\frac{2\lambda^2}{\sqrt{1+2\chi}} (\partial_\+\phi)^2 S_{=,=}\ , \label{eq: f in component} 
    \end{split}\\
    \begin{split}
	\mathbf g&\,=\,\frac{\sqrt{1+2\chi}+1}{2}-\lambda\left(\frac{1+\chi+\sqrt{1+2\chi}}{2\sqrt{1+2\chi}}\right) S_{\+,=}\cr
    &\quad-\frac{2\lambda^2}{\sqrt{1+2\chi}} (\partial_=\phi)^2 S_{\+,\+}\ , \label{eq: g in component}
    \end{split}
\end{align}
in a way that the superaction~\eqref{eq: superaction form} with the deformed superfield~\eqref{eq: modified superfield ansatz} exactly reproduces the \ttb-deformed Lagrangian after the auxiliary field $F$ is integrated out. Noting that $\mathbf f$ and $\mathbf g$ is multiplied to $\psi_+$ and $\psi_-$ respectively in \eqref{eq: modified superfield ansatz}, one can express the deformed superfield~\eqref{eq: modified superfield ansatz} compactly in term of the \ttb-deformed Lagrangian \eqref{eq: susy lag}:
%
%
\begin{equation}
	\Phi=\phi+i\theta^+\big(1-\lambda \cL\big)\psi_++i\theta^-\big(1-\lambda \cL\big)\psi_-+i\theta^+\theta^-F\ . \label{eq: modified superfield}
\end{equation}

The form of superaction~\eqref{eq: superaction form} with the vanilla superfield was observed in~\cite{Ivanov:2000nk} as a Goldstone superﬁeld action in the broken $\cN=(2,2)$ supersymmetry to $\cN= (1,1)$. However, we found that the superfield~\eqref{eq: superaction form} with the vanilla superfield~\eqref{eq: usual superfield} leads to higher derivative fermions which have larger Hilbert space than one expects. On the other hand, the superaction~\eqref{eq: superaction form} with the deformed superfield~\eqref{eq: modified superfield} does not generate higher derivatives and therefore does not give rise to the extra degrees of freedom along the \ttb \ deformation as expected. Note that the deformed superfield is related to the vanilla superfield 
\begin{equation}
	\Phi\,=\,\phi+i\theta^+\eta_++i\theta^-\eta_-+i\theta^+\theta^-F\ , \label{eq: eta superfield}
\end{equation}
by a field redefinition of the fermions
\begin{equation}
    \eta_\pm\,=\,\left(1-\lambda\cL[\phi,\psi]\right)\psi_\pm\ . \label{eq: eta def}
\end{equation}
The field redefinition accompanies the Jacobian in the path integral, which can eliminate the additional degrees of freedom in the superaction with the vanilla superfield. In Section~\ref{sec: field redefinition}, we elaborate on the elimination of the extra degrees of freedom under the field redefinition with a toy model.

Our superaction~\eqref{eq: superaction form} with the vanilla superfield~\eqref{eq: usual superfield} does not satisfy the flow equation of the SUSY \ttb \ deformation in Refs.~\cite{Baggio:2018rpv,Chang:2018dge}. The Noether procedure for supercurrent in the superspace used in Ref.~\cite{Chang:2018dge} has an ambiguity in choosing total derivative terms. It is highly interesting to find total derivative terms in the Noether procedure for the supercurrent which does not produce the higher derivative fermions under the deformation. We expect that there would exist an improved supercurrent of which the SUSY \ttb \ deformation leads to our superaction in Eq.~\eqref{eq: superaction form}.

Now let us consider the off-shell supersymmetry transformation. From the vanilla superfield, one can easily read off the off-shell supersymmetry transformation of $\phi,\eta_\pm$ and $F$:
\begin{align}
    &\delta_{susy} \phi \,=\,  -i\epsilon^+ \eta_+-i\epsilon^-\eta_-\ , \label{eq: offshell phi susy var}\\
    &\delta_{susy} \eta_+ \,=\,  2\epsilon^+\partial_\+\phi-\epsilon^- F\ , \label{eq: offshell psip susy var}\\
    &\delta_{susy} \eta_- \,=\, 2\epsilon^-\partial_=\phi+\epsilon^+ F\ , \label{eq: offshell psim susy var}\\
    &\delta_{susy} F \,=\, -2i \epsilon^+\partial_\+ \eta_- +2i \epsilon^-\partial_= \eta_+ \ . 
\end{align}
From these transformations, one can obtain the off-shell supersymmetry transformation of $\phi, \psi_\pm$. For this, we need to invert the relation~\eqref{eq: eta def} to solve for $\psi_\pm$. We find that
\begin{equation}
    \psi_\pm\,=\,\left(1-\lambda\cL[\phi,\eta]\right)^{-1}\eta_\pm\ , \label{eq: inverse eta def}
\end{equation}
where $\cL$ is the \ttb-deformed Lagrangian~\eqref{eq: susy lag} with $\psi_\pm$ replaced by $\eta_\pm$. Using Eq.~\eqref{eq: inverse eta def}, the off-shell supersymemtry transformation of $\phi,\psi_\pm$ and $F$ is found to be
\begin{align}
    \begin{split}
    \delta_{susy} \phi &\,=\,  -i\epsilon^+ \left(1-\lambda\cL[\phi,\psi]\right)\psi_+\cr
    &\quad-i\epsilon^-\left(1-\lambda\cL[\phi,\psi]\right)\psi_-\ , \label{eq: offshell susy variation of phi}
    \end{split}\\
    \delta_{susy} \psi_\pm &\,=\,  \delta_{susy}\left[\left(1-\lambda\cL[\phi,\eta]\right)^{-1}\eta_\pm\right] , \label{eq: offshell susy variation of psi from eta}\\
    \begin{split}
    \delta_{susy} F &\,=\, -2i \epsilon^+\partial_\+ \big[(1-\lambda\cL[\phi,\psi])\psi_-\big]\cr
    &\quad+2i \epsilon^-\partial_= \big[ (1-\lambda\cL[\phi,\psi])\psi_+\big] \ , \label{eq: offshell susy variation of F}
    \end{split}
\end{align}
where the transformation of $\psi_\pm$ can be evaluated by using that of $\eta_\pm$ in Eqs.~\eqref{eq: offshell psip susy var}, \eqref{eq: offshell psim susy var} and $\phi$ in Eq.~\eqref{eq: offshell susy variation of phi} as well as the relation between $\eta_\pm$ and $\psi_\pm$~\eqref{eq: eta def}.

In the Lagrangian evaluated from the superaction~\eqref{eq: superaction form} with the deformed superfield \eqref{eq: modified superfield}, one can obtain the classical equation of motion for the auxiliary field $F$. We find that the solution for the auxiliary field $F$ is given by
\begin{equation}
\begin{aligned}
    F&\,=\,{4i\lambda \over 1-8\lambda\partial_\+\phi\partial_=\phi}\;\psi_+\psi_-\big[\left(1-6\lambda\partial_\+\phi\partial_=\phi\right)\partial_\+\partial_=\phi\\
    &\quad+\lambda\left((\partial_\+\phi)^2\partial_=^2\phi+(\partial_=\phi)^2\partial_\+^2\phi\right)\big]\\
    &\quad+2i\lambda(\partial_\+\phi\; \psi_+\partial_=\psi_-+\partial_=\phi\;\partial_\+\psi_+\psi_-)\ . \label{eq: auxiliary solution}
\end{aligned}
\end{equation}
As mentioned, one can reproduce the \ttb-deformed Lagrangian~\eqref{eq: susy lag} after inserting the solution~\eqref{eq: auxiliary solution} into the Lagrangian from the superaction. Moreover, we insert the solution for $F$ \eqref{eq: auxiliary solution} into the off-shell supersymmetry transformation of $\phi$~\eqref{eq: offshell susy variation of phi} and $\psi_\pm$~\eqref{eq: offshell susy variation of psi from eta}. Then, one can express the transformations in terms of the canonical variables $\phi, \pi$ and $\psi_\pm$. We confirm that the resulting transformation agrees with the on-shell supersymmetry transformation~\eqref{eq: canonical phi susy var}$\sim$\eqref{eq: canonical psi susy var2} in the canonical analysis as expected.
\begin{alignat}{2}
    &\delta_{susy} \phi &&\,=\, i\epsilon^+ \{Q^1_+,\phi\}_\dirac + i\epsilon^- \{Q^1_-,\phi\}_\dirac\ , \label{eq: onshell susy variation 1} \\
    &\delta_{susy} \psi_\pm &&\,=\,  i\epsilon^+ \{Q^1_+,\psi_\pm\}_\dirac + i\epsilon^- \{Q^1_-,\psi_\pm\}_\dirac\ .\label{eq: onshell susy variation 2}
\end{alignat}

In~\cite{Ivanov:2000nk}, the superaction~\eqref{eq: superaction ansatz} with the vanilla superfield was obtained from the broken supersymmetry given by
\begin{equation}
\begin{aligned}
    \delta_{\broken}\Phi\,=\, -{4\pi i\over L}\big[&\epsilon_+(\theta^+-i\lambda \cD_-\cA)\\
    &+\epsilon_-(\theta^-+i\lambda \cD_+\cA)\big]\ . \label{eq: broken global susy superfield}
\end{aligned}
\end{equation}
Note that in small $\lambda$ limit this is nothing but the shifting symmetry of the fermions:
\begin{equation}
\begin{aligned}
    \left.\left(\delta_{\broken}\Phi\right)\right\vert_{\lambda\rightarrow0}&\,=\,{4\pi i\over L}\left[\theta^+\epsilon_++\theta^-\epsilon_-\right] \\
    \Rightarrow\quad \delta_{\broken}\psi_\pm&\,=\,{4\pi \over L}\epsilon_\pm\ ,
\end{aligned}
\end{equation}
where $L$ denotes the circumference of the spatial coordinate. For non-zero $\lambda$, we evaluate the on-shell non-linearly realized supersymmetry of $\phi$ and $\psi_\pm$ from Eq.~\eqref{eq: broken global susy superfield} with the deformed superfield by using the solution for $F$~\eqref{eq: auxiliary solution}. We find that this broken supersymmetry is identical to the global fermi symmetry in Eqs.~\eqref{eq: canonical phi broken susy var}$\sim$\eqref{eq: canonical psi broken susy var2} generated by $Q^2_\mp$
\begin{align}
    &\delta_{\broken} \phi = i\epsilon_+ \{Q^{\broken}_-,\phi\}_\dirac + i\epsilon_- \{Q^{\broken}_+,\phi\}_\dirac\ ,\\
    &\delta_{\broken} \psi_\pm =  i\epsilon_+ \{Q^{\broken}_-,\psi_\pm\}_\dirac + i\epsilon_- \{Q^{\broken}_+,\psi_\pm\}_\dirac\ , 
\end{align}
where the charge $Q^\broken_\pm$ for the non-linearly realized supersymmetry is identified with the supercharge $Q^2_\mp$. \ie
\begin{align}
    Q^{\broken}_\pm \,=\, Q^2_\mp\ .
\end{align}
Note that the spinor index of $Q^2_\mp$ comes from the 3D target space of $\cN=2$ GS superstring~\cite{Lee:2021iut}. And this is the reason for the discrepancy in the indices in $Q^{\broken}_\pm$ and $Q^2_\mp$. From the point of view of 3D $\cN=2$ GS superstring  action related to the \ttb \ deformation~\cite{Lee:2021iut}, the non-linearly realized symmetry generated by $Q^2_\mp$ stems from the compactified target space (\eg the compactified $X^-$ target coordinate for the discrete light-cone quantization). For the connection between the GS action and the \ttb \ deformation, the circumference of the compactified light-cone coordinate $X^-$ is proportional to the circumference $L$ of the worldsheet spatial coordinate which controls the non-linearly realized symmetry~\eqref{eq: broken global susy superfield}. In Section~\ref{sec: broken susy}, we will elaborate further on the relation between the \ttb \ deformation and the symmetry breaking.

\subsection{Field Redefinition and Path Integral Measure}
\label{sec: field redefinition}

In the path integral, one can freely perform a field redefinition, and physics should not depend on this field redefinition. This is called the equivalence theorem in quantum field theory~\cite{Umezawa:1961igm,Kamefuchi:1961sb,Salam:1970fso}. The field redefinition leads to the Jacobian in the path integral. For a field redefinition of a bosonic field $\Phi$ into $\phi$, one can exponentiate the Jacobian $\det\! \big({\delta \Phi \over \delta \phi}\big)$ by using the ghost field $b$ and $\bar{b}$:
\begin{equation}
\begin{aligned}
	Z\,=\,&\int \cD \Phi \; e^{-S[\Phi]}\,=\,\int \cD \phi \;\det \left({\delta \Phi \over \delta \phi}\right)\; e^{i S\left(\Phi[\phi]\right)}\ ,\\
	\,=\,&\int \cD \phi \cD \bar{b}\cD b  \; e^{i S\left(\Phi[\phi]\right) + i \int d^d x\; \bar{b}\left({\delta \Phi \over \delta \phi}\right) b}\ .
\end{aligned}
\end{equation}
It was shown~\cite{Slavnov:1990nd,Bastianelli:1990ey,Alfaro:1992np} that the the total action
\begin{align}
    S_{tot}\,\equiv \, S\big(\Phi[\phi]\big)+ \int d^d x\; \bar{b}{\delta \Phi \over \delta \phi} b
\end{align}
has BRST symmetry given by
\begin{align}
	\delta \phi \quad&\longrightarrow \quad \epsilon \; b\ ,\\
	\delta b\quad &\longrightarrow \quad   0\ ,\\
	\delta \bar{b} \quad &\longrightarrow \quad - \epsilon \;{\delta S \over \delta \Phi }\ .
\end{align}
Note that this BRST symmetry originates from redundant description under the arbitrary field redefinition.

The field redefinition of fermion also results in the Jacobian in the path integral. Unlike the bosonic case, the Jacobian appears in the denominator of the path integral measure, and it can be exponentiated by complex scalar fields $\gamma$ and $\bar{\gamma}$:
\begin{align}
	Z=&\int \cD \psi \; e^{iS[\psi]}=\int {\cD \eta\over \det \left({\delta \psi \over \delta \eta}\right)}  e^{iS\left(\psi[\eta]\right)}\ ,\\
	=&\int \cD \eta \cD \bar{\gamma} \cD \gamma  \; e^{i S\left(\psi[\eta]\right) + i \int d^d x\; \bar{\gamma}\left({\overleftarrow{\delta} \psi  \over \overleftarrow{\delta} \eta }\right) \gamma  }\ .
\end{align}
The total action with the exponentiated Jacobian also enjoys BRST symmetry given by
\begin{align}
	\delta \psi \quad&\longrightarrow \quad \epsilon \; \gamma \ ,\\
	\delta \gamma\quad &\longrightarrow \quad 0\ ,\\
	\delta \bar{\gamma} \quad &\longrightarrow \quad  \epsilon\;  { \overleftarrow{\delta} S \over \overleftarrow{\delta} \psi }\ .
\end{align}

It is often believed that one do not have to take the Jacobian of the path integral into account at the classical level because it does not have any influence on the classical analysis. However, the Jacobian from the field redefinition can play a crucial role even at the classical level, and one should take it into account. Especially, the field redefinition involved with the time derivatives can bring about higher derivative kinetic theory. And if it were not for the Jacobian, it would have larger Hilbert space.

To see this, let us consider the one-dimensional free massive fermion $\psi$ and $\bar{\psi}$.
\begin{align}
	L\,=\, - i \dot{\bar{\psi}} \psi + m \bar{\psi} \psi\ . \label{eq: lag free feremion}
\end{align}
Under the field redefinition of $\psi, \bar{\psi}$
\begin{align}
	\psi \,=\, \eta + i \lambda \dot{\eta}\quad,\qquad  \bar{\psi} \,=\, \bar{\eta} \ ,
\end{align}
the Lagrangian~\eqref{eq: lag free feremion} together with the exponentiated Jacobian becomes 
\begin{align}
	L_{tot}\,=\, i\lambda m \bar{\eta} \dot{\eta} - i \dot{\bar{\eta}} \eta + \lambda \dot{\bar{\eta}} \dot{\eta} + m \bar{\eta} \eta + \bar{\gamma} (1+ i \lambda \partial )\gamma \ , \label{eq: lag free fermion redefined}
\end{align}
where $\gamma$ and $\bar{\gamma}$ are complex bosons. The new Lagrangian~\eqref{eq: lag free fermion redefined} has higher derivative term compared to that of free fermion in Eq.~\eqref{eq: lag free feremion}. Such ``higher derivative'' fermionic theories have larger Hilbert space with respect to the ordinary one~\cite{Kausch:1995py,Gaberdiel:1998ps,Kausch:2000fu,LeClair:2007iy,Robinson:2009xm,Lee:2021iut,Ryu:2022ffg}. However according to the equivalence theorem in quantum field theory~\cite{Umezawa:1961igm,Kamefuchi:1961sb,Salam:1970fso}, the physical degree of freedom should not be enlarged under the field redefinition. This problem of the enlarged Hilbert space is resolved by BRST cohomology involved with the Jacobian~\cite{Slavnov:1990nd}. The BRST symmetry of the Lagrangian~\eqref{eq: lag free fermion redefined}, which originates from the field redefinition, is given by
\begin{align}
	\delta \eta  \,=\,& \epsilon\; \gamma \ ,\\
	\delta \bar{\eta}  \,=\,& 0\ , \\
	\delta \gamma\,=\,& 0\ ,\\
	\delta \bar{\gamma} \,=\,& \epsilon\;\big(-i\dot{\bar{\eta}} + m \bar{\eta}\big)\ .
\end{align}
where $\epsilon$ is a Grassmann-odd parameter. And one can evaluate the corresponding BRST charge $Q$
\begin{align}
	Q\,=\,-\big(i  \lambda m \bar{\eta}  +\lambda \dot{\bar{\eta}} \big)\gamma\ . \label{eq: brst charge q}
\end{align}

Moreover, the Lagrangian in Eq.~\eqref{eq: lag free fermion redefined} is also invariant under the following fermionic transformation
\begin{align}
	\delta \eta  \,=\,& 0 \ ,\\
	\delta \bar{\eta}  \,=\,& \bar{\epsilon}\; \bar{\gamma}\ , \\
	\delta \gamma\,=\,&-\bar{\epsilon}\;\big(-i\dot{\eta} - m \eta\big) \ ,\\
	\delta \bar{\gamma} \,=\,& 0\ ,
\end{align}
where $\bar{\epsilon}$ is a Grassmann-odd parameter. The corresponding charge $Q^\dag$ is found to be
\begin{align}
    Q^\dag\,=\, \bar{\gamma} \big( -\lambda \dot{\eta} + i m\lambda \eta \big)\ .\label{eq: brst charge qdag}
\end{align}
It turns out that $Q^\dag$ is also the BRST charge corresponding to the field redefinition
\begin{align}
	\psi \,=\, \eta \ ,\qquad  \bar{\psi} \,=\, \bar{\eta} - i \lambda \dot{\bar{\eta}} \ .
\end{align}
Note that this field redefinition also transforms the Lagrangian of the free fermion in Eq.~\eqref{eq: lag free feremion} to the higher derivative Lagrangian in Eq.~\eqref{eq: lag free fermion redefined} up to total derivative.

To analyze the physical Hilbert space, we will quantize the system in the Hamiltonian formalism. The conjugate momentum $\pi$ (and $\bar{\pi},\Pi,\bar{\Pi}$) of $\psi$ (and $\bar{\psi},\gamma,\bar{\gamma}$, respectively) is found to be
\begin{alignat}{3}
	\pi \,=\,&  i \lambda m \bar{\eta} + \lambda \dot{\bar{\eta}} \  , \qquad &&\bar{\pi} \,=\,-i\eta  + \lambda \dot{\eta}\ ,\label{eq: conj mom fermion}\\
	\Pi\,=\, & i \lambda \bar{\gamma}\  ,\qquad  &&\bar{\Pi}\,=\,  0\ .\label{eq: conj mom boson}
\end{alignat}
For the case of ordinary fermion, $\dot{\eta}$ and $\dot{\bar{\eta}}$ do not appear in the Eq.~\eqref{eq: conj mom fermion}, and those equations form the second class constraints. These constraints halve the dimension of the Hilbert space. But, for our case, the time derivative of $\eta$ and $\bar{\eta}$ can be expressed in terms of other canonical variables, and Eq.~\eqref{eq: conj mom fermion} does not play a role of the constraint anymore. 
%
%
This implies that the Hilbert space is doubled compared to the ordinary fermionic theory. On the other hand, Eq.~\eqref{eq: conj mom boson} does not contain the time derivative of $\eta$ and $\bar{\eta}$. Hence they form the second class constraints of the system:
\begin{align}
    \cC_1\,=\, \Pi -i\lambda \bar{\gamma}\ ,\quad \cC_2\,=\, \bar{\Pi}\ ,
\end{align}
From these Dirac brackets, one can obtain the Dirac bracket of boson $\gamma$ and $\bar{\gamma}$
\begin{align}
		\{\gamma, \bar{\gamma}\}^-_{\text{\tiny D}}\,=\, - {i\over \lambda} \ .
\end{align}
and the other non-vanishing Dirac brackets are
\begin{align}
    \{\eta,\pi\}^+_{\text{\tiny D}}\,=\, 1\ ,\qquad \{\bar{\eta},\bar{\pi}\}^+_{\text{\tiny D}}\,=\, -1\ .
\end{align}
After promoting the Dirac bracket to the canonical (anti-)commutation relation, we have
\begin{align}
		\{\eta, \pi\}\,=\,i \ ,\qquad \{ \bar{\eta},\bar{\pi}\} \,=\,-i\ ,\qquad [\gamma,\bar{\gamma}]\,=\, {1\over \lambda} \ .\label{eq: canonical comm rel}
\end{align}
Now we will express those operators in terms of the fermi oscillator $b,b^\dag, c,c^\dag$ and bosonic oscillator $a, a^\dag$ satisfying the following (anti-)commutation relation.
\begin{align}
    \{b,b^\dag\}\,=\,1\ ,\qquad \{c,c^\dag\}\,=\,-1\ ,\qquad [a,a^\dag]\,=\, 1\ .
\end{align}
Note that the right hand side of the anti-commutation relation $\{c,c^\dag\}$ has the opposite sign to that of the ordinary fermi oscillator. This fermi oscillator with the unusual anti-commutation relation has been observed in the quantization of higher derivative fermionic theories~\cite{LeClair:2007iy,Robinson:2009xm,Lee:2021iut,Ryu:2022ffg}. And this unusual anti-commutation relation leads to the negative norm state with respect to the naive inner product.\footnote{Here, we assume that the vacuum has a positive norm.} \ie
\begin{align}
    \big\Vert c^\dag |0\rangle \big\Vert^2\,=\,\langle 0 | c c^\dag |0\rangle \,=\, - \langle 0 | 0 \rangle\ .
\end{align}
To avoid this negative norm, a new inner product $\langle\; \cdot\; \rangle_\cJ$ by inserting an operator $\cJ$ into the naive inner product defined by
\begin{align}
    \langle \cO \rangle_\cJ \equiv \langle \cJ \, \cO\rangle\hspace{10mm}\mbox{where}\quad \cJ\equiv \exp (i\pi c^\dag c)\label{def: cj norm}
\end{align}
was proposed~\cite{LeClair:2007iy,Robinson:2009xm,Lee:2021iut}. With respect to this $\cJ$-inner product, the state $c^\dag |0\rangle$ does have a positive norm.
\begin{align}
    \big\Vert c^\dag |0\rangle \big\Vert_\cJ^2 \,=\, \langle 0 | c \cJ c^\dag |0\rangle \,=\, \langle 0 | 0\rangle \ .
\end{align}
Furthermore, it was proved~\cite{Lee:2021iut} that the path integral formalism of this type of model is consistent with the operator formalism with $\cJ$-inner product. Accordingly, a bra state should be defined with respect to the $\cJ$-inner product. And in transforming a ket state to a bra state, one has to use the $\cJ$-Hermitian adjoint $\dag_\cJ$ on the operator, which is defined by
\begin{align}
    \cO^{\dag_\cJ}\,\equiv \, \cJ \cO^\dag \cJ\ .
\end{align}

Coming back to the quantization, we find the transformation parametrized by $\theta\in \mathbb{R}$ from $\eta,\bar{\eta},\pi$ and $\bar{\pi}$ to $b,b^\dag,c$ and $c^\dag$ 
%
%
%
\begin{align}
	\bar{\pi} + i \eta\,=\, & i (1+\lambda m)^{1\over 2} (\sinh \theta b + \cosh \theta c) \ ,\\
	\pi -  im\lambda \bar{\eta}\,=\, & -i (1+\lambda m)^{1\over 2}(\sinh \theta b^\dag + \cosh \theta c^\dag) \ , \\
	\bar{\pi} -  im\lambda\eta\,=\, &  -i (1+\lambda m)^{1\over 2}(\cosh \theta b + \sinh \theta c)\ ,\\
	\pi + i \bar{\eta}\,=\, &  i (1+\lambda m)^{1\over 2} (\cosh \theta b^\dag + \sinh \theta c^\dag)\ .
\end{align}
The parameter $\theta$ is involved with Bogoliubov transformation~\cite{Lee:2021iut} among $b,b^\dag,c$ and $c^\dag$, and therefore those oscillators (and their vacuum) implicitly depend\footnote{Therefore, strictly speaking, we should have explicitly indicated the $\theta$-dependence of the oscillators and vacuum. For example, $b_\theta,b^\dag_\theta,$ and $|\theta\rangle$. However, for simplicity, we omit the $theta$ symbol  in the oscillators and the vacuum in this paper.} on the parameter $\theta$.

The bosonic oscillators $a,a^\dag$ are identified with $\gamma, \bar{\gamma}$ by
\begin{alignat}{4}
    a\,=\, & \sqrt{\lambda}\gamma\quad,\quad &&a^\dag\,=\,&  \sqrt{\lambda} \bar{\gamma} \hspace{10mm} &&\mbox{for} \quad \lambda>0\ ,\\
    a\,=\, & \sqrt{|\lambda|}\bar{\gamma}\quad,\quad &&a^\dag\,=\, &  \sqrt{|\lambda|} \gamma\hspace{10mm} &&\mbox{for} \quad \lambda<0\ .
\end{alignat}
One can obtain the Hamiltonian form the Lagrangian~\eqref{eq: lag free fermion redefined}, and it can be expressed in terms of the oscillators $a,b,c$:
\begin{equation}
\begin{aligned}
	H&\,=\, \pi \dot{\eta} + \dot{\bar{\eta}} \bar{\pi} + \Pi \dot{\gamma} +\bar{\Pi}\dot{\bar{\gamma}}- L\ ,\cr
	&\,=\, {1\over \lambda} \big( \pi -   im\lambda\bar{\eta} \big) \big( \bar{\pi} + i \eta \big)\cr
    &\quad- m \bar{\eta}\eta +\dot{\gamma} \big(\Pi -i \lambda \bar{\gamma}\big) +\bar{\Pi}\dot{\bar{\gamma}} - \bar{\gamma}\gamma\ ,\cr
    &\,=\, \bigg({1+\lambda m\over \lambda}\sinh^2\theta-{me^{2\theta}\over 1+m\lambda} \bigg)b^\dag b\cr
    &\quad+\bigg({1+\lambda m\over \lambda}\cosh^2\theta -{me^{2\theta}\over 1+m\lambda} \bigg) c^\dag c - {1\over \lambda}a^\dag a\cr
    &\quad+\bigg({1+\lambda m \over 2\lambda}\sinh2\theta -{me^{2\theta}\over 1+m\lambda}\bigg)(b^\dag c+ c^\dag b)\ .
\end{aligned}
\end{equation}
Here, we consider the case where $\lambda>0$ and $0<m\lambda<1$. Furthermore, the BRST charges in Eq.~\eqref{eq: brst charge q} and Eq.~\eqref{eq: brst charge qdag} can also be written as
\begin{align}
    \begin{split}
    Q&\,=\,- \pi \gamma\cr
    &\,=\, -{i\over \big[\lambda(1+ \lambda m)\big]^{1\over 2}}\bigg[ \big(m\lambda \cosh \theta - \sinh\theta \big)b^\dag\cr
    &\quad+ \big(m\lambda \sinh \theta - \cosh\theta \big)c^\dag \bigg] a\ \ ,
    \end{split}\\
    \begin{split}
    Q^\dag&\,=\, -  \bar{\gamma}\big[\bar{\pi}+i(1-m\lambda)\eta\big] \ , \cr 
    &\,=\, {i\over \big[\lambda(1+ \lambda m)\big]^{1\over 2}} a^\dag \bigg[ \big(m\lambda \cosh \theta - \sinh\theta \big)b\cr
    &\quad+ \big(m\lambda \sinh \theta - \cosh\theta \big) \bigg]\ \ .
    \end{split}
\end{align}
For our purpose, it is convenient to choose a particular value of $\theta$  
\begin{align}
	\tanh\theta \,=\, \begin{cases}
     \;\lambda m &\quad\mbox{for}\quad |m\lambda|<1\ , \\
     \; {1\over\lambda m}&\quad\mbox{for}\quad |m\lambda|>1\ ,
    \end{cases}
\end{align}
where the cross term $b^\dag c + c^\dag b$ in the Hamiltonian vanishes. For this special value of $\theta$, the Hamiltonian and the BRST charges become
\begin{align}
	H\,=\,& -m b^\dag b + {1\over \lambda } c^\dag c - {1\over \lambda }a^\dag a\ \label{eq: simple ham},\\
    Q\,=\,& i \bigg({1-m\lambda\over\lambda} \bigg)^{1\over 2}c^\dag a\ ,\\
    Q^\dag\,=\,& -i \bigg({1-m\lambda\over\lambda} \bigg)^{1\over 2} a^\dag c\ .
\end{align}
The energy spectrum is not bounded below because of the bosonic oscillator. This unbounded energy spectrum will be resolved after demanding the invariance under the BRST charges. However note that the Hamiltonian can be written in terms of $Q, Q^\dag$ as
\begin{align}
    H\,=\, -mb^\dag b + {1\over 1-m\lambda }\{Q^\dag,Q\} \ .
\end{align}
Usually, this seemingly implies that the Hamiltonian would be bounded below\footnote{Here we consider the case where $m\lambda<1$. For other cases, one can also get a similar result.}, which contradicts the unbounded spectrum which can be seen in Eq.~\eqref{eq: simple ham}. As mentioned in Eq.~\eqref{def: cj norm}, one should use the $\cJ$-inner product which is consistent with the path integral formalism. Therefore, noting that
\begin{align}
    \big(Q^{\dag}\big)^{\dag_\cJ}\,=\, \cJ Q\cJ \,=\, -Q\ ,
\end{align}
one can show the $\{Q^\dag,Q\}$ is non-positive with respect to $\cJ$-norm. \ie
\begin{align}
    \langle \Psi| \{ Q^\dag,Q\} |\Psi\rangle_\cJ\,=\, - \big\Vert Q |\Psi\rangle \big\Vert^2_\cJ \ . 
\end{align}

Among the Fock states
\begin{equation}
\begin{aligned}
    |n_b,n_c;n_a\rangle \,\equiv \, {1\over \sqrt{n_a!}}\big(b^\dag\big)^{n_b}\big(c^\dag\big)^{n_c}\big(a^\dag\big)^{n_a}|0\rangle\\
    (n_b,n_c=0,1 \;\;\mbox{and}\;\; n_a= 0,1,2,\cdots) \ ,
\end{aligned}
\end{equation}
it is easy to see that only two states, $|0,0;0\rangle$ and $|1,0;0\rangle$, are annihilated by $Q$ and $Q^\dag$, and they exactly reproduce the spectrum of the original free fermion.\footnote{Instead of demanding that the physical states are annihilated by $Q$ and $Q^\dag$, one may consider the BRST cohomology, which leads to the same conclusion.}
\begin{align}
    H |0,0;0\rangle \,=\, 0\quad,\qquad H|1,0;0\rangle \,=\, -m|1,0;0\rangle\ .
\end{align}
Therefore, although a field redefinition can change a model into the higher derivative one which has larger Hilbert space, the path integral measure from the field redefinition plays a crucial role in eliminating the spurious degrees of freedom to reproduce the same physical results as those of the original model.

\section{Broken supersymmetry}
\label{sec: broken susy}

In this section, we identify the \ttb-deformed theories as symmetry-broken models. We have already seen in Section~\ref{sec: Construction of superaction with deformed superfield} that the deformed $\cN=(1,1)$ free theory contains not only the $\cN=(1,1)$ supercharge $Q^1_\pm$ but also the fermi global charge $Q^2_\pm$ that corresponds to the broken supersymmetry. The \ttb-deformed scalar field and fermion, which are described by Nambu-Goto action at static gauge \cite{Cavaglia:2016oda,Conti:2018jho} and the Volkov-Akulov~(VA) model~\cite{Cribiori:2019xzp} respectively, is also an example of the symmetry broken model~\cite{Nielsen:1973cs,Volkov:1973ix}. In~\cite{Jiang:2019hux,Chang:2018dge, Cribiori:2019xzp, Chang:2019kiu, Ferko:2019oyv}, the SUSY breaking has been also studied for the theories deformed by the manifestly supersymmetric \ttb \ deformation operator.

In this work, we will use two approaches to construct the symmetry-broken models: the non-linear realization method and the constrained superfield method. And we identify these results with the \ttb-deformed theories. In Section~\ref{sec: nonlinear}, we have used the non-linear realization with the spacetime symmetry breaking. Likewise, we will construct the simplest spontaneously symmetry-broken models respecting the spacetime symmetry-breaking pattern by following the recipes of non-linear realization~\cite{Volkov:1973ix,Volkov:1973vd,Ogievetsky:1974,Penco:2020kvy}. The models built from the non-linear realization turn out to be off-shell equivalent to the \ttb-deformed theories. We will show that the \ttb \ deformation of free scalar, fermion and $\cN=(1,1)$ theory are identified as the actions from non-linear realization with the various spacetime symmetry-breaking patterns. In Section~\ref{sec: constrained superfield}, we will use the constrained superfield method to construct the SUSY broken model, such as the VA model~\cite{Rocek:1978nb,Komargodski:2009rz}. Also, the constrained superfield method makes the model with PBGS as manifestly supersymmetric superaction form~\cite{Bagger:1996wp,Ivanov:2000nk}. For example, the superaction of the constrained superfield for the spontaneously broken SUSY to $\cN=(1,1)$ \cite{Ivanov:2000nk} is of the same form as that in Eq.~\eqref{eq: superaction form}. And the superactions from more generic SUSY breaking patterns by the constrained superfield method is identified with SUSY \ttb-deformed results \cite{Ferko:2019oyv,Jiang:2019hux}. But as we have emphasized, one needs the field redefinition of fermions \eqref{eq: eta def} to identify the action \eqref{eq: superaction of PBGS} from the constrained superfield with the original \ttb-deformed theory \eqref{eq: superaction form}.

\subsection{Non-linear realization}
\label{sec: nonlinear}

A non-linear realization is a powerful tool for studying spontaneously symmetry-broken models in physics. It provides a systematic way to construct effective field theories~(EFT) for spontaneously broken symmetries, including both internal and spacetime symmetries. The idea of the non-linear realization was first developed by Callan, Coleman, Wess, and Zumino in the context of the chiral symmetry breaking in QCD~\cite{Coleman:1969sm,Callan:1969sn} and it has since been extended to include the breaking of spacetime symmetries~\cite{Volkov:1973ix,Volkov:1973vd}.

In the non-linear realization, the Goldstone fields are treated as coordinates on a coset space, which is defined as the quotient space of the original symmetry group $G$ by its unbroken subgroup $H$. The generators of the unbroken subgroup act linearly on this space, while those of the broken subgroup act non-linearly. This leads to a non-linear transformation law for the Goldstone fields. The (left-invariant) Maurer-Cartan form $g^{-1}\mathrm{d}g$ with coset element $g$ is efficient to construct the action that is invariant under these non-linearly transformed Goldstone fields because the Maurer-Cartan form is invariant under the global transformation of $G$ by construction. Moreover, one can extract the components that covariantly transform under the local transformation of $H$. By extracting these components from the Maurer-Cartan form, one can construct the minimal effective theory that respects the symmetry-breaking pattern. More details can be found in~\cite{Ogievetsky:1974,Penco:2020kvy}. To get an idea of how to obtain the action from the coset element and its Maurer-Cartan form, we start with the simple examples which will be identified as \ttb \ deformation of free scalar and fermion theory. Then we apply the non-linear realization method to get the \ttb-deformed free $\cN=(1,1)$ theory~\eqref{eq: susy lag}.

\vspace{5mm}

\noindent \textbf{Example \textrm{I}}\hspace{5mm} We will consider the effective theory with the symmetry breaking pattern---the three-dimensional Poincare symmetry group $G$ broken down to the two-dimensional one $H$. The generators of the 3D Poincare algebra are denoted by $P_i$ and $J_{ij}$ $(i,j=0,1,2)$. This 3D Poincare symmetry is broken to the 2D Poincare symmetry of which generators are denoted by $P_a$ and $J_{ab}$ ($a=0,1$). And the generators of the broken part are $P_2$ and $K_a\equiv J_{a2}$.

Now let us consider the following representation of the coset element $g\in G/H$.
\begin{equation}
    g\,=\,e^{x\cdot P}e^{\tilde\phi P_2}e^{v\cdot K}\ ,\label{eq: coset repre}
\end{equation}
where $V\cdot W\equiv \eta_{ab}V^aW^b$. As mentioned, the coordinates $\tilde\phi(x)$ and $v^a(x)$ on a coset space will act as Goldstone fields. Note the coset representation~\eqref{eq: coset repre} of $g$ is changed under the left transformation by $g_0$
\begin{align}
    g(x,\tilde\phi,v)=e^{x\cdot P}e^{\tilde\phi P_2}e^{v\cdot K}\;\;\rightarrow \;\; g_0\, g= g' \, h(\tilde{\phi},v,g_0)\ ,\label{eq: left transf of coset element}
\end{align}
where $g'\equiv g(x',\tilde{\phi}',v')$ and $h(\tilde{\phi},v,g_0)\in H$. This can be seen as the field-dependent non-linear transformation of the Goldstone fields $\tilde{\phi}$ and $v$.

The Maurer-Cartan form is given by
\begin{equation}
\begin{aligned}
    &g^{-1}\mathrm{d}g\\
    &=\,e^{-v\cdot K}\left(dx\cdot P+d\tilde\phi P_2 \right) e^{v\cdot K}+ e^{-v\cdot K}de^{v\cdot K}\cr
    &=\,dx^\mu E^a_{\ \mu} P_a  + \text{other generators} \label{eq: mc form ex1} \ ,
\end{aligned}
\end{equation}
where we evaluate $E^a_{\ \mu}$ explicitly in Appendix~\ref{app: coset construction calculation}:
\begin{equation}
    \begin{aligned}
    E^a_{\ \mu}&\,=\, \delta^a_{\ \mu}+\left(\cosh(\sqrt{-v\cdot v})-1\right)\frac{v^a  v_\mu}{v\cdot v}\\
    &\quad-\sinh(\sqrt{-v\cdot v})\frac{v^a\partial_\mu\tilde\phi }{\sqrt{-v\cdot v}}\ .
    \end{aligned}
\end{equation}
Under the left transformation of the coset element, the Maurer-Cartan form is transformed as
\begin{align}
    g^{-1}dg\;\;\rightarrow\;\; {g'}^{-1}dg'= h g^{-1}dg h + h^{-1}dh\ .
\end{align}
Under this transformation, ${E^a}_\mu$ in Eq.~\eqref{eq: mc form ex1} is transformed like ``vielbein'' with local Lorentz transformation depending on Goldstone fields.\footnote{Henceforth, as we only work on the two-dimensional model, we will call such object as coset zweibein.} For more details, see \cite{Ogievetsky:1974,Penco:2020kvy}. This is one of the building blocks to construct the invariant action under the left global transformation of the coset element by $g_0\in G$. For example, the measure $d^2x\; \det E^a_{\ \mu}$ is invariant under the local transformation, which gives us one minimal way to construct the invariant action. By introducing the dimensionful parameter $\lambda$ which would be identified with the \ttb \ deformation parameter, we have an action which respects the symmetry-breaking pattern in this example:
\begin{equation}
    \begin{aligned}
    S&\,=\,-\frac1{2\lambda}\int d^2x\; \det(E^a_{\ \mu})\\
    &\,=\,-\frac1{2\lambda}\int d^2x\; \bigg( \cosh(\sqrt{-v\cdot v})\\
    &\quad+\sinh(\sqrt{-v\cdot v})\frac{v\cdot \partial\tilde\phi}{\sqrt{-v\cdot v}}\bigg)\ .
    \end{aligned}
\end{equation}
Since $v^a(x)$ is non-dynamical, one can integrate it out by using its equation of motion\footnote{Or, one can obtain the same result by the ``inverse Higgs mechanism'' \cite{Ivanov:1975zq}.}
\begin{equation}
\partial_a \tilde{\phi}\,=\,\tanh (\sqrt{-v\cdot v})\frac{v_a }{\sqrt{-v\cdot v}} \label{eq: nondynamical v eq} \ .
\end{equation}
Then the Lagrangian becomes
\begin{equation}
    \cL\,=\,-\frac1{2\lambda} \sqrt{1+\partial\tilde\phi\cdot\partial\tilde\phi}\,=\,-\frac1{2\lambda} \sqrt{1-8\lambda\partial_\+\phi\partial_=\phi} \ ,
\end{equation}
where we rescaled the dimensionful Goldstone field $\tilde\phi(x)=\sqrt{2\lambda}\phi(x)$ by the dimensionful parameter $\lambda$ to have dimensionless Goldstone field. Up to constant, this action is exactly the same as the one produced by \ttb \ deformation of free scalar theory \cite{Cavaglia:2016oda,Conti:2018jho}
\begin{equation}
    \cL_{T\bar T} \,=\,-\frac1{2\lambda} \left(\sqrt{1-8\lambda\partial_\+\phi\partial_=\phi}-1\right)\ .
\end{equation}

\vspace{5mm}

\noindent \textbf{Example \textrm{II}}\hspace{5mm} We can also apply the formalism of the non-linear realization method to the superspace symmetry breaking. In this example, we consider the symmetry breaking pattern---2D \texorpdfstring{$\cN=(1,1)$}{N=1,1} super-Poincare group $G$ to 2D Poincare group $H$. It turns out that the minimal construction of such effective action gives the well-known VA model \cite{Volkov:1973ix} which is well-known for the spontaneously supersymmetry broken model. Here the generators for the unbroken group $H$ are the two dimensional translation $P_a$ and the Lorentz rotation $J_{ab}$ ($a,b=0,1$) while $Q_\alpha$ for $\cN=(1,1)$ supersymmetry is broken. Thus, we can start with the coset element by attaching Goldstino fields $\theta^\alpha(x)$ to the broken supersymmetry generator $Q_\alpha$ as
\begin{equation}
    g\,=\,e^{x\cdot P}e^{\theta^\alpha Q_\alpha}\ .
\end{equation}
The non-linear transformation of the Goldstino field $\theta^\alpha$ under the broken supersymmetry can be easily shown by acting the constant group element to the coset element. From
\begin{equation}
    e^{\epsilon^\alpha Q_\alpha}e^{x^aP_a+\theta^\beta Q_\beta}\,=\,e^{(x^a+i\bar\epsilon\gamma^{a}\theta(x)){P_a}}e^{(\theta^\alpha(x)+\epsilon^\alpha){Q_\alpha}}\ ,
\end{equation}
we get
\begin{equation}
    \delta \theta_\alpha(x)=\epsilon_\alpha-i \bar{\epsilon} \gamma^a \theta(x) \partial_a \theta_\alpha(x)\ . \label{VA field variation}
\end{equation}
To construct the effective model, we begin with the Maurer-Cartan form
\begin{equation}
    g^{-1}\mathrm{d}g \,=\,d x^\mu e^a_{\ \mu} P_a+ d x^\mu \partial_\mu \theta^\alpha Q_\alpha\ ,
\end{equation}
The coset zweibein $e^a_{\ \mu}$ is found to be\footnote{See Appendix~\ref{app: coset construction calculation} for details.}
\begin{equation}
    e^a_{\ \mu}\,=\,\delta_{\ \mu}^a-i \bar{\theta} \gamma^a \partial_\mu \theta\ . \label{eq: VA coset zweibein }
\end{equation}
As before, the simplest effective model for spontaneously broken supersymmetry can be written as
\begin{equation}
\begin{aligned}
\cL&\,=\, -\frac{1}{\lambda} \det(e^a_{\ \mu})\\
&\,=\,  -\frac{1}{\lambda} \bigg(1-2i\theta_+\partial_=\theta_+-2i\theta_-\partial_\+\theta_-\\
&\quad+4\theta_+\partial_\+\theta_+\theta_-\partial_=\theta_--4\theta_+\partial_=\theta_+\theta_-\partial_\+\theta_-\bigg)\ .
\end{aligned}\label{eq: nonlinear realization VA action}
\end{equation}
By rescaling the Goldstino fields into the two-dimensional Majorana fermion fields $\theta_\pm=\sqrt{\lambda/2}\,\psi_\pm$, we have two dimensional VA model
\begin{equation}
    \cL\,=\, S_{\+,=} +  S_{=,\+} + \lambda( S_{\+,\+} S_{=,=}-S_{\+,=}S_{=,\+})-\frac{1}{\lambda}\ , \label{eq: VA lag}
\end{equation}
which is equivalent to the \ttb-deformed free fermion theory up to constant.
\begin{equation}
    \cL_{T\bar T}\,=\,S_{\+,=} +  S_{=,\+} + \lambda( S_{\+,\+} S_{=,=}-S_{\+,=}S_{=,\+})\ .\label{eq: deformed free fermion lag}
\end{equation}
The Goldstino fields $\psi_\alpha$'s non-linear field variations~\eqref{VA field variation} now become
\begin{equation}
\begin{aligned}
    \delta\psi_+\,=\,&\tilde\epsilon_+\Big[1-\lambda S_{\+,=}\Big]-i\tilde\epsilon_-\Big[\lambda\psi_-\partial_\+\psi_+\Big]\ ,\\
    \delta\psi_-\,=\,&-i\tilde\epsilon_+\Big[\lambda\psi_+\partial_=\psi_-\Big]+\tilde\epsilon_-\Big[1-\lambda S_{=,\+}\Big]\ ,  \label{eq: TT fermion field variation}
\end{aligned}
\end{equation}
where $\tilde\epsilon_\alpha=\sqrt{2/\lambda}\epsilon_\alpha$.

\vspace{5mm}

\noindent \textbf{3D \texorpdfstring{$\cN=2$}{N=2} super-Poincare symmetry to 2D \texorpdfstring{$\cN=(1,1)$}{N=1,1} super-Poincare symmetry}\hspace{5mm} Our main interest, \ttb-deformed $\cN=(1,1)$ free theory, has the unbroken $\cN=(1,1)$ supersymmetry and the broken supersymmetry which can be incorporated in 3D $\cN=2$ super-Poincare algebra with topological charge~\cite{Lee:2021iut}. Thus, for the construction of the effective theory with the same symmetry-breaking pattern via the non-linear realization method, let us consider the following coset element.
\begin{equation}
    g\,=\, e^{x\cdot P}e^{\tilde\phi(x) P_2}e^{\theta^\alpha(x)Q_\alpha}e^{v(x)\cdot K}\ , 
\end{equation}
where $P_a$ ($a=0,1$) is the unbroken generator while $Q_\pm, P_2$ and $K_a$ corresponds to broken one. We can extract the coset zweibein $\mathbb E^a_{\ \mu}$ from the Maurer-Cartan form
\begin{equation}
\begin{aligned}
     &g^{-1}\mathrm{d}g\cr
     &=e^{-v\cdot K}\left((dx-i\bar\theta\gamma d\theta)\cdot P+d\tilde\phi P_2 +d\theta^\alpha Q_\alpha \right) e^{v\cdot K}\cr
     &\quad+ e^{-v\cdot K}de^{v\cdot K}\cr
     &= dx^\mu\mathbb E^a_{\ \mu} P_a+ \text{other generators} \ ,
\end{aligned}
\end{equation}
where
\begin{equation}
\begin{aligned}
    \mathbb E^a_{\ \mu}&\,=\, (\delta^b_{\ \mu}-i\bar\theta\gamma^b \partial_\mu\theta)\\
    &\quad\times\left(\delta^a_{\ b}+\left(\cosh(\sqrt{-v\cdot v})-1\right)\frac{v^a v_b  }{v\cdot v}\right)\cr
    &\quad+\sinh(\sqrt{-v\cdot v})\frac{v^a \partial_\mu\tilde\phi}{\sqrt{-v\cdot v}}\cr
    &\,=\, e^b_{\ \mu}\bigg(\delta^a_{\ b}+\left(\cosh(\sqrt{-v\cdot v})-1\right)\frac{ v^a v_b}{v\cdot v}\cr
    &\quad+\sinh(\sqrt{-v\cdot v})\frac{v^a \nabla_b\tilde\phi}{\sqrt{-v\cdot v}} \bigg)\ .
\end{aligned}
\end{equation}
where $\nabla_b\equiv (e^{-1})_b^{\ \mu} \partial_\mu$ and $e^a_{\ \mu}=\delta_{\ \mu}^a-i \bar{\theta} \gamma^a \partial_\mu \theta$. Note that $e^b_{\ \mu}$ is the coset zweibein which appeared in the VA model~\eqref{eq: VA coset zweibein }. Taking the determinant on the coset zweibein ${\mathbb{E}^a}_\mu$, one can obtain the effective action which respects the symmetry-breaking pattern:
\begin{equation}
\begin{aligned}
    S&\,=\,-\frac1{2\lambda}\int d^2x\; \det(\mathbb E^a_{\ \mu})\\
    &\,=\,-\frac1{2\lambda}\int d^2x\;\bigg[ \det(e^a_{\ \mu})\\
    &\quad\times\left( \cosh(\sqrt{-v\cdot v})+\sinh(\sqrt{-v\cdot v})\frac{v\cdot \nabla\tilde\phi}{\sqrt{-v\cdot v}}\right)\bigg]\ .
\end{aligned}
\end{equation}
As in the first example, we can integrate out the non-dynamical field $v^a(x)$ to get the effective action: 
\begin{align}
    S\,=\,-\frac1{2\lambda}\int d^2x\; \det(e^a_{\ \mu})\sqrt{1+\nabla\tilde\phi\cdot\nabla\tilde\phi}\ . \label{eq: Nonlinearly realized (1,1) supersymmetric theory 1}
\end{align}
One may take Eq.~\eqref{eq: Nonlinearly realized (1,1) supersymmetric theory 1} as an effective action respecting the required symmetry-breaking pattern. However, we find out that we need to add the VA action~\eqref{eq: nonlinear realization VA action} to Eq.~\eqref{eq: Nonlinearly realized (1,1) supersymmetric theory 1} in order to contact with the \ttb-deformed $\cN=(1,1)$ theory. After we rescale the Goldstone fields as $\tilde\phi=\sqrt{2\lambda}\phi\,,\,\theta_\pm=\sqrt{\lambda/2}\,\psi_\pm$ and adding some constant term, we have
\begin{align}
    \begin{aligned}
    \cL&=\frac1{2\lambda}\left(2-\det(\mathbb E^a_{\ \mu})-\det(e^a_{\ \mu})\right)\\
    &=\frac{1}{\lambda}\left(1-\frac{1}{2}\det(e^a_{\ \mu})\left(1+\sqrt{1-8\lambda\nabla_\+\phi\nabla_=\phi}\right)\right)\ ,\label{eq: Nonlinearly realized (1,1) supersymmetric theory 2}
    \end{aligned}
\end{align}
where
\begin{align}
    \begin{split}
    \nabla_\+\phi&\,=\,\partial_\+\phi+\lambda(\partial_=\phi S_{\+,\+}+\partial_\+\phi S_{=,\+})\cr
    &\quad+\lambda^2 S_{\+,\+}(\partial_=\phi S_{=,\+}+\partial_\+\phi S_{=,=})\ ,
    \end{split}\\
    \begin{split}
    \nabla_=\phi&\,=\,\partial_=\phi+\lambda(\partial_\+\phi S_{=,=}+\partial_=\phi S_{\+,=})\cr
    &\quad+\lambda^2 S_{=,=}(\partial_\+\phi S_{\+,=}+\partial_=\phi S_{\+,\+})\ .
    \end{split}
\end{align}
Indeed, one can see that this action~\eqref{eq: Nonlinearly realized (1,1) supersymmetric theory 2} coincides with Lagrangian density~\eqref{eq: susy lag} of the \ttb-deformed $\cN=(1,1)$ theory. Moreover, the VA action that we added is turned out to correspond to the Wess-Zumino~(WZ) term\footnote{See the term $\cB$ in Eq.~(4.8) and Eq.~(4.10)~of Ref.~\cite{Lee:2021iut}.} in the GS-like action in static gauge~\cite{Lee:2021iut}. Therefore, we expect that the WZ term in the coset superspace~\cite{Gomis:2006wu} would reproduce the VA action that we added. We leave it for future works.

\subsection{Constrained superfield}
\label{sec: constrained superfield}

For the spontaneous supersymmetry breaking, there is another method to construct the effective action for Goldstone fields which is called the constrained superfield method~\cite{Rocek:1978nb,Komargodski:2009rz}. This method was initiated in Ref.~\cite{Rocek:1978nb} where the VA model was studied with the constrained superfield. This approach has an advantage in constructing the models in manifestly supersymmetric form with superfields and superaction.

It turned out that some models generated by the constrained superfield method are non-trivially related to the theories constructed from the non-linear realization method via the field redefinition~\cite{Zheltukhin:2010xr,Kuzenko:2010ef,Kuzenko:2011tj}.

We also provide the relations between the supersymmetry broken models in Section~\ref{sec: nonlinear} and the models constructed by the constrained superfield method. As we have pointed out in Sec~\ref{sec: Construction of superaction with deformed superfield}, the superaction~\eqref{eq: superaction form} for the \ttb-deformed $\cN=(1,1)$ free theory can indeed identified with the one constructed by the constrained superfield method~\cite{Ivanov:2000nk} with the vanilla scalar superfield~\eqref{eq: usual superfield} via the field redefinitions~\eqref{eq: eta def}.

\subsubsection{\texorpdfstring{\ttb}{TTbar} deformed fermion theory from \texorpdfstring{$\cN=(1,1)$}{N=1,1} constrained superfield}


The superaction of a single $\cN=(1,1)$ scalar superfield $\Phi$ can be written with the general potential term $V(\Phi)$ as
\begin{equation}
    \frac12 D_-\Phi D_+\Phi-V(\Phi)\ .
\end{equation}
With the nilpotent superfield constraint~\cite{Komargodski:2009rz}
\begin{equation}
    \Phi^2\,=\,0\ ,
\end{equation}
the potential term $V(\Phi)$ gets truncated to linear $\Phi$ term. Then, the Lagrangian density can be written as
\begin{align}
\cL_{KS}\,=\,&\int d^2\theta\ \left(\frac12 D_-\Phi D_+\Phi-i\alpha\Phi\right)\ . \label{eq: KS lagrangian}
\end{align}
where $\alpha$ is a real constant. The constraint and the equation of motion reads
\begin{equation}
\Phi^2\,=\,0\ ,\quad \alpha\Phi-i\Phi D_-D_+\Phi	\,=\,0\ . \label{eq: eqs in constrained (1,1) superfield}
\end{equation}
One can solve these equations for the components $\phi, F$ of the $\cN=(1,1)$ scalar superfield
\begin{equation}
    \Phi\,=\,\phi + i\theta^+\eta_+ + i\theta^-\eta_- + i\theta^+\theta^- F \ ,
\end{equation}
in terms of the fermion component fields $\eta_\pm$ as
\begin{align}
	&\phi\,=\,\frac{i\eta_+\eta_-}{F}\ , \label{eq: KS phi sol}\\
    \begin{split}
	&F\,=\,\alpha -{2i\over\alpha}(\eta_+\partial_=\eta_+ +\eta_-\partial_\+\eta_-)\cr
    &\qquad+{4\over\alpha^3}(\eta_+\partial_\+\eta_+ \eta_-\partial_=\eta_-+3\eta_+\partial_=\eta_+ \eta_-\partial_\+\eta_-)\ . \label{eq: KS F sol}
    \end{split}
\end{align}
After plugging back these solutions  into the Lagrangian density~\eqref{eq: KS lagrangian}, we have
\begin{equation}
    \begin{aligned}
\cL_{KS}&\,=\, i\eta_+\partial_=\eta_+ +i\eta_-\partial_\+\eta_--\frac{2}{\alpha^{2}}\big(\eta_+\partial_\+\eta_+ \eta_-\partial_=\eta_-\\
&\quad+3\eta_+\partial_=\eta_+ \eta_-\partial_\+\eta_-\big)-{\alpha^2\over2}\ .\label{eq: KS lag eta only}
    \end{aligned}
\end{equation}
In addition, from the solution~\eqref{eq: KS phi sol} and \eqref{eq: KS F sol}, one can find the supersymmetric variation of $\eta_\pm$:
\begin{align}
    \begin{split}
    &\delta \eta_+ \,=\,  2i\alpha^{-1}\epsilon^+\left[\eta_+\partial_\+\eta_--\eta_-\partial_\+\eta_+\right]\cr
    &\qquad-\alpha\epsilon^-\Big[1-2i\alpha^{-2}(\eta_+\partial_=\eta_+ +\eta_-\partial_\+\eta_-)\cr
    &\qquad+4\alpha^{-4}\left(\eta_+\partial_\+\eta_+ \eta_-\partial_=\eta_-+3\eta_+\partial_=\eta_+ \eta_-\partial_\+\eta_-\right)\Big]\ , \label{eq: KS etap susy var}
    \end{split}
    \\
    \begin{split}
    &\delta \eta_- \,=\,2i\alpha^{-1}\epsilon^-\left[\eta_+\partial_=\eta_--\eta_-\partial_=\eta_+\right]\cr
    &\qquad+\alpha\epsilon^+\Big[1-2i\alpha^{-2}(\eta_+\partial_=\eta_+ +\eta_-\partial_\+\eta_-)\cr
    &\qquad+4\alpha^{-4}\left(\eta_+\partial_\+\eta_+ \eta_-\partial_=\eta_-+3\eta_+\partial_=\eta_+ \eta_-\partial_\+\eta_-\right)\Big]\ .\label{eq: KS etam susy var}
    \end{split}
\end{align}
It turns out that under the field redefinition
\begin{align}
    \eta_+&\quad\longrightarrow\quad (1-2\alpha^{-2}S_{=,\+})\psi_+\ ,\\
    \eta_-&\quad\longrightarrow\quad (1-2\alpha^{-2}S_{\+,=})\psi_-\ ,
\end{align}
where $S_{=,\+},S_{\+,=}$ are fermion bilinears of $\psi_\pm$~\eqref{eq: fermion bilinears}, the Lagrangian density $\cL_{KS}$~\eqref{eq: KS lag eta only} can be written in terms of $\psi_\pm$ as follows
\begin{align}
    \cL_{KS}= S_{\+,=} + S_{=,\+} +\lambda\left(S_{\+,\+}S_{=,=}-S_{\+,=}S_{=,\+}\right)-\frac1\lambda\; .\label{eq: KS lag psi}
\end{align}
Here, we identified $\lambda=2\alpha^{-2}$ and $\tilde\epsilon_\pm = \mp\alpha\epsilon^\mp$. The Lagrangian density~\eqref{eq: KS lag psi} is identical to that of the VA model~\eqref{eq: VA lag} as well as the \ttb-deformed free fermion theory~\eqref{eq: deformed free fermion lag} up to constant term. Furthermore, the transformation of $\psi_\pm$ in Eq.~\eqref{eq: KS etap susy var} and Eq.~\eqref{eq: KS etam susy var} reads
\begin{align}
    \delta\psi_+\,=\,&\tilde\epsilon_+\Big[1-\lambda S_{\+,=}\Big]-i\tilde\epsilon_-\Big[\lambda\psi_-\partial_\+\psi_+\Big]\ , \label{eq: KS psi transf 1} \\
    \delta\psi_-\,=\,&-i\tilde\epsilon_+\Big[\lambda\psi_+\partial_=\psi_-\Big]+\tilde\epsilon_-\Big[1-\lambda S_{=,\+}\Big]\ .\label{eq: KS psi transf 2}
\end{align}
This transformation $\psi_\pm$ also agrees with that of VA model~\eqref{eq: TT fermion field variation}.

\subsubsection{\texorpdfstring{\ttb}{TTbar} deformed \texorpdfstring{$\cN=(1,1)$}{N=1,1} theory from \texorpdfstring{$\cN=(2,2)$}{N=2,2} constrained superfield}


The superactions of the Goldstone superfields with various supersymmetry breaking patterns were studied in Ref.~\cite{Ivanov:2000nk} where the constrained superfield method was used to produce such models with partially broken global supersymmetry~(PBGS). We find that the supraction from the constrained superfield method for the broken $\cN=(2,2)$ SUSY to $\cN=(1,1)$ is identical to the superaction that we constructed in Eq.~\eqref{eq: superaction form} with $\cN=(1,1)$ vanilla scalar superfield~\eqref{eq: usual superfield} replaced by the deformed superfield~\eqref{eq: modified superfield}. We review the constrained superfield method in Ref.~\cite{Ivanov:2000nk} to build the effective superaction for the supersymmetry breaking pattern $\cN=(2,2)$ to $\cN=(1,1)$.

We begin with the $\cN=(2,2)$ chiral superfield $\boldsymbol{\Phi}$ with the chirality condition
\begin{align}
    \bar D_\pm\boldsymbol{\Phi}\,=\,0\ ,
\end{align}
This condition enables us to write the chiral superfield $\boldsymbol{\Phi}$ in terms of two $\cN=(1,1)$ real superfields $\boldsymbol{\varphi},\boldsymbol{\phi}$ as
\begin{align}
        \begin{split}
        \boldsymbol{\Phi} \,=\,& (\boldsymbol{\varphi}+i\boldsymbol{\phi})\\
        &+i\theta^+_2(i\cD_+\boldsymbol{\varphi}-\cD_+\boldsymbol{\phi})+i\theta^-_2(i\cD_-\boldsymbol{\varphi}-\cD_-\boldsymbol{\phi})\cr 
        &+i\theta^+_2\theta^-_2(-i\cD_-\cD_+\boldsymbol{\varphi}+\cD_-\cD_+\boldsymbol{\phi})\ ,
        \end{split}\\
        \begin{split}
        \bar{\boldsymbol{\Phi}} \,=\,& (\boldsymbol{\varphi}-i\boldsymbol{\phi})\\
        &+i\theta^+_2(-i\cD_+\boldsymbol{\varphi}-\cD_+\boldsymbol{\phi})+i\theta^-_2(-i\cD_-\boldsymbol{\varphi}-\cD_-\boldsymbol{\phi})\cr 
        &+i\theta^+_2\theta^-_2(-i\cD_-\cD_+\boldsymbol{\varphi}-\cD_-\cD_+\boldsymbol{\phi})\ .
        \end{split}
\end{align}
Then, one can express the superaction for the free $\cN=(2,2)$ chiral superfield $\boldsymbol{\Phi}$ in terms of $\boldsymbol{\varphi},\boldsymbol{\phi}$:
\begin{equation}
\begin{aligned}
    S&\,=\,{1\over2}\int d^2x\, d\vartheta^+d\vartheta^-d\bar{\vartheta}^-d\bar{\vartheta}^+\ \bar{\boldsymbol{\Phi}}\boldsymbol{\Phi}\\
    &\,=\, \frac{1}{2}\int d^2x\, d^2\theta_1 \left(\cD_-\boldsymbol{\varphi}\cD_+\boldsymbol{\varphi} + \cD_-\boldsymbol{\phi}\cD_+\boldsymbol{\phi}  \right)\ , \label{eq: free 2,2 superaction}
\end{aligned}
\end{equation}
where $d^2\theta_{1,2}=d\theta^+_{1,2}d\theta^-_{1,2}$ and we dropped the total derivative term. From the $\cN=(2,2)$ chiral superfield, one can construct the $\cN=(2,2)$ real scalar superfield $\boldsymbol{\Phi}_{real}$ with a constant $v\in \mathbb{R}$ of dimension 1:
\begin{equation}
\begin{aligned}
    &\boldsymbol{\Phi}_{real}\,\equiv\, {1\over2}(\bar{\boldsymbol{\Phi}}+\boldsymbol{\Phi})+iv\theta^+_2\theta^-_2\\
    &=\,\boldsymbol{\varphi}-i\theta^+_2\cD_+\boldsymbol{\phi}-i\theta^-_2\cD_-\boldsymbol{\phi}+\theta^+_2\theta^-_2\left(iv+\cD_-\cD_+\boldsymbol{\varphi}\right)\ .
\end{aligned}
\end{equation}
Note that $v$ plays a role of the vacuum expectation value. Now we impose the nilpotent constraint on the $\cN=(2,2)$ real superfield as in the previous section: 
\begin{align}
    \boldsymbol{\Phi}_{real}^2\,=\,0 \ . \label{eq: nilpotency 2,2}
\end{align}
From the expansion of the constraint~\eqref{eq: nilpotency 2,2} with respect to $\theta_2^\pm$, we have
\begin{gather}
    \boldsymbol{\varphi}^2\,=\,0\ ,\quad\boldsymbol{\varphi} \cD_\pm \boldsymbol{\phi}\,=\,0\ ,  \label{eq: 2,2 contraint components12}\\
    i v\boldsymbol{\varphi}+\boldsymbol{\varphi}\cD_-\cD_+\boldsymbol{\varphi}-\cD_-\boldsymbol{\phi}\cD_+\boldsymbol{\phi}\,=\,0 \ . \label{eq: 2,2 contraint components3}
\end{gather}
One can take the following ansatz for $\boldsymbol{\varphi}$ which trivially solve the constraint~\eqref{eq: 2,2 contraint components12}
\begin{align}
    \boldsymbol{\varphi}\,=\,\Omega[\boldsymbol{\phi}]\, \cD_-\boldsymbol{\phi}\cD_+\boldsymbol{\phi}\ .
\end{align}
The remaining constraint~\eqref{eq: 2,2 contraint components3} can be written as
\begin{equation}
    i v\Omega + \Omega^2 \cD_-\cD_+(\cD_-\boldsymbol{\phi}\cD_+\boldsymbol{\phi})-1\,=\,0 \ .
\end{equation}
Among the two solutions $\Omega_\pm$ of the quadratic equation, we choose $\Omega_+$, and the superfield $\boldsymbol{\varphi}$ is found to be
\begin{equation}
    \begin{aligned}
    \boldsymbol{\varphi}&\,=\, \Omega_+[\boldsymbol{\phi}]\, \cD_-\boldsymbol{\phi}\cD_+\boldsymbol{\phi}\\
    &\,=\,{-2i v^{-1} \cD_-\boldsymbol{\phi}\cD_+\boldsymbol{\phi} \over 1+\sqrt{1-4v^{-2}\cD_-\cD_+(\cD_-\boldsymbol{\phi}\cD_+\boldsymbol{\phi})}}\ . \label{eq: varphi solution}
    \end{aligned}
\end{equation}
To express the superaction~\eqref{eq: free 2,2 superaction} in terms of $\boldsymbol{\varphi}$, it is convenient to use the constraints~\eqref{eq: 2,2 contraint components12} $\sim$\eqref{eq: 2,2 contraint components3} to obtain the following identity.
\begin{equation}
\begin{aligned}
    0&\,=\,\frac{1}{2}\cD_-\cD_+(\boldsymbol{\varphi}^2)\,=\,\cD_-\boldsymbol{\varphi}\cD_+\boldsymbol{\varphi}+\boldsymbol{\varphi}\cD_-\cD_+\boldsymbol{\varphi}\\
    &\,=\,\cD_-\boldsymbol{\varphi}\cD_+\boldsymbol{\varphi} + \cD_-\boldsymbol{\phi}\cD_+\boldsymbol{\phi}-iv\boldsymbol{\varphi}\ , \label{eq: eq from 2,2 nilpotency}
\end{aligned}
\end{equation}
From this identity, one can express the $\cN=(2,2)$ superaction~\eqref{eq: free 2,2 superaction} in terms of $\boldsymbol{\varphi}$, and in turn, in terms of $\boldsymbol{\phi}$ by using the solution~\eqref{eq: varphi solution}:
\begin{equation}
\begin{aligned}
     S&=\frac{1}{2}\int d^2x\, d^2\theta_1 \left(\cD_-\boldsymbol{\varphi}\cD_+\boldsymbol{\varphi} + \cD_-\boldsymbol{\phi}\cD_+\boldsymbol{\phi}  \right)\\
     &=  {iv\over 2}\int d^2x\, d^2\theta_1\ \boldsymbol{\varphi} \ ,\\
     &=  \int d^2x\, d^2\theta_1  { \cD_-\boldsymbol{\phi}\cD_+\boldsymbol{\phi} \over 1+\sqrt{1-4v^{-2}\cD_-\cD_+(\cD_-\boldsymbol{\phi}\cD_+\boldsymbol{\phi})}}\ .\label{eq: superaction of PBGS}
\end{aligned}
\end{equation}
This is the superaction with PBGS pattern $\cN=(2,2)$ to $\cN=(1,1)$ in terms of the $\cN=(1,1)$ real Goldstone superfield $\boldsymbol{\phi}$. The superaction form~\eqref{eq: superaction of PBGS} itself is exactly the same as the one we constructed in Eq.~\eqref{eq: superaction form}, once we identify $v$ to \ttb \ deformation parameter as $\lambda=2v^{-2}$ and replace the $\cN=(1,1)$ vanilla superfield $\boldsymbol{\phi}$ by the modified superfield $\Phi$~\eqref{eq: modified superfield}.

\section{Conclusion}
\label{sec: conclusion}

In this work, we have constructed the superaction with the deformed superfield for the \ttb-deformed $\cN=(1,1)$ supersymmetric model, and we have studied the \ttb-deformed theory from the point of view of the effective action for the symmetry breaking. We reviewed the deformed $\cN=(1,1)$ on-shell supersymmetry and the global fermi symmetry in the \ttb-deformed free $\cN=(1,1)$ SUSY model~\cite{Lee:2021iut}, and these global symmetries can be identified with the broken super-Poincare symmetry for 3D target space. We have shown that one cannot form the superaction without higher-derivative fermion for the \ttb-deformed free $\cN=(1,1)$ SUSY model from the vanilla superfield. Thus, instead of the vanilla superfield, we constructed the superaction with the deformed superfield. We obtained the \ttb-deformed off-shell $\cN=(1,1)$ supersymmetry, and we proved that this deformed off-shell supersymmetry leads to the known deformed on-shell $\cN=(1,1)$ supersymmetry~\cite{Lee:2021iut} after integrating out the auxiliary field. Furthermore, we found that this superaction is related to the effective superaction for the Goldstone superfield in the broken supersymmetry. We demonstrated that the non-linearly realized off-shell supersymmetry becomes the fermi global on-shell symmetry~\cite{Lee:2021iut} after the auxiliary field is integrated out. The deformed superfield is related to the vanilla superfield under the field redefinition, which leads to a non-trivial path integral measure. With a toy model, we demonstrated that the path integral measure from the field redefinition eliminates the additional inflow of the degrees of freedom in the canonical analysis.

We studied the \ttb \ deformation in terms of the symmetry broken models. From the method of the non-linear realization, we reproduced the \ttb-deformed free scalar, fermions, and $\cN=(1,1)$ supersymmetric theory. Furthermore, we reviewed the constrained superfield method which provides the superaction form of \ttb \ deformation of the free fermions and $\cN=(1,1)$ supersymmetric theory. This method has been used to study the SUSY \ttb-deformed theory as symmetry broken models \cite{Ferko:2019oyv,Jiang:2019hux}. We showed that the theory from the constrained superfield method can exactly be identified with the original \ttb-deformed theory after the field redefinition. 

The correspondence of the effective models constructed from the non-linear realization and the constrained superfield method up to field redefinition has been widely investigated in various models~\cite{Zheltukhin:2010xr,Kuzenko:2010ef,Kuzenko:2011tj}. Assuming that the relation between the \ttb \ deformation and the symmetry broken models persist in the higher-dimensional models, it would be interesting to investigate the higher-dimensional effective theory for the symmetry breaking as the higher-dimensional version of the \ttb \ deformation. Most investigations have been limited to the classical level. It remains to be investigated whether this symmetry breaking pattern is also realized at quantum level~\cite{Dubovsky:2012sh}.

We have shown that the field redefinition of the fermion in the deformed superfield~\eqref{eq: modified superfield}, or equivalently, the path integral measure for the vanilla superfield is inevitable in the superaction~\eqref{eq: superaction form} for the \ttb \ deformation of the free $\cN=(1,1)$ SUSY model in order to avoid the emergence of the unexpected degree of freedom. However, the origin of this field redefinition or the path integral measure is still not clear at this moment. Moreover, our deformed superfield might not be the unique resolution to this issue. Since the \ttb \ deformation of general 2D QFT is known to be equivalent to the QFT coupled to the 2D flat Jackiw-Teitelboim~(JT) gravity~\cite{Dubovsky:2017cnj,Dubovsky:2018bmo,He:2022bbb}, it would be interesting to understand the origin of the path integral measure by investigating the 2D flat JT gravity coupled to the $\cN=(1,1)$ SUSY model.

The path integral in terms of the configuration variables also has ambiguity in the path integral measure. We have shown that the \ttb-deformed $\cN=(1,1)$ SUSY action is invariant under the off-shell~\eqref{eq: offshell susy variation of phi}$\sim$\eqref{eq: offshell susy variation of F} and the on-shell~\eqref{eq: onshell susy variation 1}$\sim$\eqref{eq: onshell susy variation 2} supersymmetry transformation, which are highly non-linear transformations of the configuration variables. Hence, although the action is invariant, it is not clear whether the path integral measure is invariant under such non-linear transformations or not. This non-linear symmetry might be anomalous at quantum level. Otherwise, the path integral should have non-trivial measure which is invariant under the non-linear transformation. In fact, the path integral could acquire a non-trivial contribution in the measure when we integrate out the auxiliary field $F$ in the superaction or when we integrate out the conjugate momenta from the first-order Lagrangian to obtain the second-order one. This contribution to the path integral measure is irrelevant at classical level, while the path integral measure discussed in Section~\ref{sec: field redefinition} plays a crucial role even at classical level. We hope to report on the \ttb-deformed SUSY transformation at quantum level in forthcoming work.

It is natural to extend our analysis to the \ttb \ deformation of the other two-dimensional supersymmetric models such as \ttb-deformed free $\cN=(2,2),\,(4,4)$ supersymmetric models~\cite{Frolov:2019nrr} or \ttb-deformed interacting SUSY models. The rationale behind this work can be equally applied to the deformation of generic theories with supersymmetry in terms of component fields. Though the computations are quite involved, the canonical analysis can be done to check the broken and unbroken supersymmetries. We hope that the connections with the symmetry-broken models via the constrained superfield method would provide more hints about the origin and its general rule for the field redefinition. We leave them for future work.

We would like to thank Yuji Hirono and Euihun Joung for useful discussions. Also, we thank Piljin Yi for collaboration in the early stages of this work and for valuable suggestions and discussions. K.L. was supported by Basic Science Research Program through the National Research Foundation of Korea(NRF) funded by the Ministry of Education(NRF-2020R1I1A2054376). The work of J.Y. was supported by the National Research Foundation of Korea (NRF) grant funded by the Korea government (MSIT) (No. 2019R1F1A1045971, 2022R1A2C1003182). J.Y. is supported by an appointment to the JRG Program at the APCTP through the Science and Technology Promotion Fund and Lottery Fund of the Korean Government. J.Y. is also supported by the Korean Local Governments - Gyeongsangbuk-do Province and Pohang City.

\appendix

\section{Conventions}
\label{app: convention}

\begin{itemize}
    \item $(t,x)$ : (dimensionful) coordinates for 2D space-time 
    \item $a=0,1$ : two-dimensional flat indices, $\mu=0,1$ : two-dimensional curved indices, $i=0,1,2$ : three-dimensional flat indices, $\alpha=-,+$ : Spinor indices
    \item $\dot\phi\,\equiv\, \partial_0\phi,\quad \phi^\prime\,\equiv\, \partial_1\phi $ : time and space derivative
    \item Light-cone coordinates
    \begin{align}
        	x^\pm\,\equiv\, t\pm x\ .
    \end{align}
    \begin{align}
	\partial_\+\,=\,{1\over 2}(\partial_0 + \partial_1)\;\;,\qquad \partial_= \,=\,{1\over 2}(\partial_0 - \partial_1)\ .
    \end{align}
    \item $\lambda$ : \ttb \ deformation parameter of dimension length-squared.
    \item $\phi,\psi_\pm$ and $\spi,\fpi_\pm$ : scalar field, fermion and its conjugate momentum.
    \item Two-dimensional Lorentzian metric and gamma matrices
    \begin{align}
        \eta=\left(\begin{array}{cc}
    -1 & 0 \\
    0 & 1
    \end{array}\right), \quad \gamma^{0}=\left(\begin{array}{cc}
    0 & -i \\
    i & 0
    \end{array}\right), \quad \gamma^{1}=\left(\begin{array}{ll}
    0 & i \\
    i & 0
    \end{array}\right)
    \end{align}
\end{itemize}

\paragraph{2D $\cN=(1,1)$ SUSY conventions}

\begin{itemize}
    \item $\theta^\pm$ : Grassman odd variables 
    \item $\int d^2x\,\equiv\,\int dt\, dx\, , \qquad \int d^2\theta\,\equiv\, \int d\theta^+\, d\theta^-$
    \item SUSY-covariant derivatives, SUSY operators :
    \begin{equation}
        \cD_\pm=-i\partial_\pm-2\theta^\pm\partial_{\pm\pm}\, ,\quad \cQ_\pm=-i\partial_\pm+2\theta^\pm\partial_{\pm\pm}
    \end{equation}
    \item $\Phi$ : 2D $\cN=(1,1)$ real scalar superfield
\end{itemize}

\paragraph{2D $\cN=(2,2)$ SUSY conventions}

\begin{itemize}
    \item $\vartheta^\pm\,\equiv\,\theta_1^\pm+i\theta_2^\pm\,,\quad \bar\vartheta^\pm\,\equiv\,\theta_1^\pm-i\theta_2^\pm$ : Complex Grassmann odd variables
    \item SUSY-covariant derivatives, SUSY operators :
    \begin{align}
        D_{\pm} &\,=\,\frac{\partial}{\partial \vartheta^{\pm}}-i \bar{\vartheta}^{\pm} \partial_{\ppmm}\ , \quad \bar{D}_{\pm} \,=\,-\frac{\partial}{\partial \bar{\vartheta}^{\pm}}+i \vartheta^{\pm} \partial_{\ppmm}
    \end{align}
    \begin{align}
        Q_{\pm} &\,=\,\frac{\partial}{\partial \vartheta^{\pm}}+i \bar{\vartheta}^{\pm} \partial_{\ppmm}\ , \quad \bar{Q}_{\pm} \,=\,-\frac{\partial}{\partial \bar{\vartheta}^{\pm}}-i \vartheta^{\pm} \partial_{\ppmm}
    \end{align}
    \item $\boldsymbol{\Phi}$ : 2D $\cN=(2,2)$ scalar superfield
\end{itemize}


\section{Useful Formulas for Non-linear Realization}
\label{app: coset construction calculation}

Here we present the useful expression for performing the non-linear realization in Sec.~\ref{sec: nonlinear}:
\begin{equation}
    \begin{aligned}
    e^{-A} d e^A&=\sum_{n=0}^{\infty} \frac{(-1)^{n}}{(n+1) !}\left(a d_{A}\right)^{n} d A\\
    &=dA-\frac12[A,dA]+\frac1{3!}[A,[A,dA]]+\cdots\ ,  \label{eq: BCH1}
    \end{aligned}
\end{equation}
\begin{equation}
    \begin{aligned}
    e^{-A}Be^A&=\sum_{n=0}^\infty\frac{(-1)^n}{n!}\left(a d_{A}\right)^{n}B\\
    &=B-[A,B]+\frac{1}{2!}[A,[A,B]]+\dots\ ,\label{eq: BCH2}
    \end{aligned}
\end{equation}
\begin{equation}
    e^Ae^B=e^{A+B+\frac{1}{2}[A, B]+\frac{1}{12}[A,[A, B]]+\frac{1}{12}[B,[B, A]]+\ldots}\ . \label{eq: BCH3}
\end{equation}
Two dimensional $\cN=(1,1)$ supersymmetry algebra satisfy
\begin{equation}
\{Q_\alpha,Q_\beta\}=-2i(\gamma^0\gamma^a)_{\alpha\beta}P_a\ , \quad [Q_\alpha, P_a]=0\ , \label{eq: 2d susy}
\end{equation}
and three-dimensional Poincare algebra satisfy
\begin{equation}
\begin{aligned}
[J_{ij},J_{kl}]&\,=\, -\eta_{ik}J_{jl}-\eta_{jl}J_{ik}+\eta_{il}J_{jk}+\eta_{jk}J_{il} \ ,\\  \quad[J_{ij},P_k]&\,=\,\eta_{ik}P_j-\eta_{jk}P_i\ . \label{eq: 3d poincare}
\end{aligned}
\end{equation}
We can use Eq.~(\ref{eq: BCH2}, \ref{eq: 3d poincare}) to have
\begin{equation}
\begin{aligned}
    e^{-v\cdot K}P_a e^{v\cdot K} &=P_a-v_a P_2-\frac1{2!}v_a (v\cdot P)\\
    &\quad+\frac{(v\cdot v)}{3!}v_aP_2-\frac{(v\cdot v)}{4!}v_a(v\cdot P)\dots\\
    &=P_a+\left(\cosh(\sqrt{-v\cdot v})-1\right)\frac{v_a(v\cdot P)}{v\cdot v}\\
    &\quad-\sinh(\sqrt{-v\cdot v})\frac{v_a P_2}{\sqrt{-v\cdot v}}\ , \label{eq: adj prod Pa}
\end{aligned}
\end{equation}
\begin{equation}
\begin{aligned}
    e^{-v\cdot K}P_2 e^{v\cdot K}&=P_2+(v\cdot P)-\frac{(v\cdot v)}{2}P_2\\
    &\quad-\frac{(v\cdot v)}{3!}(v\cdot P)+\frac{(v\cdot v)^2}{4!}P_2+\dots\\
    &\,=\,\cosh(\sqrt{-v\cdot v})P_2\\
    &\quad+\sinh(\sqrt{-v\cdot v})\frac{v\cdot P}{\sqrt{-v\cdot v}}\ . \label{eq: adj prod P2}
\end{aligned}
\end{equation}
where $K_a\equiv J_{a2}$ and $[v\cdot K,P_a]\,=\,v_a P_2$, $[v\cdot K,P_2]\,=\,-v\cdot P$.
And by using Eq.~\eqref{eq: BCH1} and Eq.~\eqref{eq: 2d susy}, we have
\begin{equation}
\begin{aligned}
    e^{-\theta^\alpha Q_\alpha}de^{\theta^\beta Q_\beta}&\,=\, d\theta^\alpha Q_\alpha+\frac12\theta^\alpha d\theta^\beta \left\{Q_\alpha, Q_\beta\right\}\\
    &\,=\, d\theta^\alpha Q_\alpha-i\bar\theta\gamma^a d\theta P_a\ . \label{eq: adj diff prod Q}
\end{aligned}
\end{equation}

\bibliography{susytt}

\begin{thebibliography}{49}%
\makeatletter
\providecommand \@ifxundefined [1]{%
 \@ifx{#1\undefined}
}%
\providecommand \@ifnum [1]{%
 \ifnum #1\expandafter \@firstoftwo
 \else \expandafter \@secondoftwo
 \fi
}%
\providecommand \@ifx [1]{%
 \ifx #1\expandafter \@firstoftwo
 \else \expandafter \@secondoftwo
 \fi
}%
\providecommand \natexlab [1]{#1}%
\providecommand \enquote  [1]{``#1''}%
\providecommand \bibnamefont  [1]{#1}%
\providecommand \bibfnamefont [1]{#1}%
\providecommand \citenamefont [1]{#1}%
\providecommand \href@noop [0]{\@secondoftwo}%
\providecommand \href [0]{\begingroup \@sanitize@url \@href}%
\providecommand \@href[1]{\@@startlink{#1}\@@href}%
\providecommand \@@href[1]{\endgroup#1\@@endlink}%
\providecommand \@sanitize@url [0]{\catcode `\\12\catcode `\$12\catcode
  `\&12\catcode `\#12\catcode `\^12\catcode `\_12\catcode `\%12\relax}%
\providecommand \@@startlink[1]{}%
\providecommand \@@endlink[0]{}%
\providecommand \url  [0]{\begingroup\@sanitize@url \@url }%
\providecommand \@url [1]{\endgroup\@href {#1}{\urlprefix }}%
\providecommand \urlprefix  [0]{URL }%
\providecommand \Eprint [0]{\href }%
\providecommand \doibase [0]{http://dx.doi.org/}%
\providecommand \selectlanguage [0]{\@gobble}%
\providecommand \bibinfo  [0]{\@secondoftwo}%
\providecommand \bibfield  [0]{\@secondoftwo}%
\providecommand \translation [1]{[#1]}%
\providecommand \BibitemOpen [0]{}%
\providecommand \bibitemStop [0]{}%
\providecommand \bibitemNoStop [0]{.\EOS\space}%
\providecommand \EOS [0]{\spacefactor3000\relax}%
\providecommand \BibitemShut  [1]{\csname bibitem#1\endcsname}%
\let\auto@bib@innerbib\@empty
\bibitem [{\citenamefont {Zamolodchikov}(2004)}]{Zamolodchikov:2004ce}%
  \BibitemOpen
  \bibfield  {author} {\bibinfo {author} {\bibfnamefont {Alexander~B.}\
  \bibnamefont {Zamolodchikov}},\ }\bibfield  {title} {\enquote {\bibinfo
  {title} {{Expectation value of composite field T anti-T in two-dimensional
  quantum field theory}},}\ }\href@noop {} {\  (\bibinfo {year} {2004})},\
  \Eprint {http://arxiv.org/abs/hep-th/0401146} {arXiv:hep-th/0401146}
  \BibitemShut {NoStop}%
\bibitem [{\citenamefont {Smirnov}\ and\ \citenamefont
  {Zamolodchikov}(2017)}]{Smirnov:2016lqw}%
  \BibitemOpen
  \bibfield  {author} {\bibinfo {author} {\bibfnamefont {F.~A.}\ \bibnamefont
  {Smirnov}}\ and\ \bibinfo {author} {\bibfnamefont {A.~B.}\ \bibnamefont
  {Zamolodchikov}},\ }\bibfield  {title} {\enquote {\bibinfo {title} {{On space
  of integrable quantum field theories}},}\ }\href {\doibase
  10.1016/j.nuclphysb.2016.12.014} {\bibfield  {journal} {\bibinfo  {journal}
  {Nucl. Phys. B}\ }\textbf {\bibinfo {volume} {915}},\ \bibinfo {pages}
  {363--383} (\bibinfo {year} {2017})},\ \Eprint
  {http://arxiv.org/abs/1608.05499} {arXiv:1608.05499 [hep-th]} \BibitemShut
  {NoStop}%
\bibitem [{\citenamefont {Cavagli\`a}\ \emph {et~al.}(2016)\citenamefont
  {Cavagli\`a}, \citenamefont {Negro}, \citenamefont {Sz\'ecs\'enyi},\ and\
  \citenamefont {Tateo}}]{Cavaglia:2016oda}%
  \BibitemOpen
  \bibfield  {author} {\bibinfo {author} {\bibfnamefont {Andrea}\ \bibnamefont
  {Cavagli\`a}}, \bibinfo {author} {\bibfnamefont {Stefano}\ \bibnamefont
  {Negro}}, \bibinfo {author} {\bibfnamefont {Istv\'an~M.}\ \bibnamefont
  {Sz\'ecs\'enyi}}, \ and\ \bibinfo {author} {\bibfnamefont {Roberto}\
  \bibnamefont {Tateo}},\ }\bibfield  {title} {\enquote {\bibinfo {title} {{$T
  \bar{T}$-deformed 2D Quantum Field Theories}},}\ }\href {\doibase
  10.1007/JHEP10(2016)112} {\bibfield  {journal} {\bibinfo  {journal} {JHEP}\
  }\textbf {\bibinfo {volume} {10}},\ \bibinfo {pages} {112} (\bibinfo {year}
  {2016})},\ \Eprint {http://arxiv.org/abs/1608.05534} {arXiv:1608.05534
  [hep-th]} \BibitemShut {NoStop}%
\bibitem [{\citenamefont {Jiang}(2021)}]{Jiang:2019epa}%
  \BibitemOpen
  \bibfield  {author} {\bibinfo {author} {\bibfnamefont {Yunfeng}\ \bibnamefont
  {Jiang}},\ }\bibfield  {title} {\enquote {\bibinfo {title} {{A pedagogical
  review on solvable irrelevant deformations of 2D quantum field theory}},}\
  }\href {\doibase 10.1088/1572-9494/abe4c9} {\bibfield  {journal} {\bibinfo
  {journal} {Commun. Theor. Phys.}\ }\textbf {\bibinfo {volume} {73}},\
  \bibinfo {pages} {057201} (\bibinfo {year} {2021})},\ \Eprint
  {http://arxiv.org/abs/1904.13376} {arXiv:1904.13376 [hep-th]} \BibitemShut
  {NoStop}%
\bibitem [{\citenamefont {Lee}\ \emph {et~al.}(2021)\citenamefont {Lee},
  \citenamefont {Yi},\ and\ \citenamefont {Yoon}}]{Lee:2021iut}%
  \BibitemOpen
  \bibfield  {author} {\bibinfo {author} {\bibfnamefont {Kyung-Sun}\
  \bibnamefont {Lee}}, \bibinfo {author} {\bibfnamefont {Piljin}\ \bibnamefont
  {Yi}}, \ and\ \bibinfo {author} {\bibfnamefont {Junggi}\ \bibnamefont
  {Yoon}},\ }\bibfield  {title} {\enquote {\bibinfo {title} {{$ T\overline{T}
  $-deformed fermionic theories revisited}},}\ }\href {\doibase
  10.1007/JHEP07(2021)217} {\bibfield  {journal} {\bibinfo  {journal} {JHEP}\
  }\textbf {\bibinfo {volume} {07}},\ \bibinfo {pages} {217} (\bibinfo {year}
  {2021})},\ \Eprint {http://arxiv.org/abs/2104.09529} {arXiv:2104.09529
  [hep-th]} \BibitemShut {NoStop}%
\bibitem [{\citenamefont {Coleman}\ \emph {et~al.}(2019)\citenamefont
  {Coleman}, \citenamefont {Aguilera-Damia}, \citenamefont {Freedman},\ and\
  \citenamefont {Soni}}]{Coleman:2019dvf}%
  \BibitemOpen
  \bibfield  {author} {\bibinfo {author} {\bibfnamefont {Evan~A.}\ \bibnamefont
  {Coleman}}, \bibinfo {author} {\bibfnamefont {Jeremias}\ \bibnamefont
  {Aguilera-Damia}}, \bibinfo {author} {\bibfnamefont {Daniel~Z.}\ \bibnamefont
  {Freedman}}, \ and\ \bibinfo {author} {\bibfnamefont {Ronak~M.}\ \bibnamefont
  {Soni}},\ }\bibfield  {title} {\enquote {\bibinfo {title} {{$ T\overline{T} $
  -deformed actions and (1,1) supersymmetry}},}\ }\href {\doibase
  10.1007/JHEP10(2019)080} {\bibfield  {journal} {\bibinfo  {journal} {JHEP}\
  }\textbf {\bibinfo {volume} {10}},\ \bibinfo {pages} {080} (\bibinfo {year}
  {2019})},\ \Eprint {http://arxiv.org/abs/1906.05439} {arXiv:1906.05439
  [hep-th]} \BibitemShut {NoStop}%
\bibitem [{\citenamefont {Baggio}\ \emph {et~al.}(2019)\citenamefont {Baggio},
  \citenamefont {Sfondrini}, \citenamefont {Tartaglino-Mazzucchelli},\ and\
  \citenamefont {Walsh}}]{Baggio:2018rpv}%
  \BibitemOpen
  \bibfield  {author} {\bibinfo {author} {\bibfnamefont {Marco}\ \bibnamefont
  {Baggio}}, \bibinfo {author} {\bibfnamefont {Alessandro}\ \bibnamefont
  {Sfondrini}}, \bibinfo {author} {\bibfnamefont {Gabriele}\ \bibnamefont
  {Tartaglino-Mazzucchelli}}, \ and\ \bibinfo {author} {\bibfnamefont
  {Harriet}\ \bibnamefont {Walsh}},\ }\bibfield  {title} {\enquote {\bibinfo
  {title} {{On $ T\overline{T} $ deformations and supersymmetry}},}\ }\href
  {\doibase 10.1007/JHEP06(2019)063} {\bibfield  {journal} {\bibinfo  {journal}
  {JHEP}\ }\textbf {\bibinfo {volume} {06}},\ \bibinfo {pages} {063} (\bibinfo
  {year} {2019})},\ \Eprint {http://arxiv.org/abs/1811.00533} {arXiv:1811.00533
  [hep-th]} \BibitemShut {NoStop}%
\bibitem [{\citenamefont {Chang}\ \emph {et~al.}(2019)\citenamefont {Chang},
  \citenamefont {Ferko},\ and\ \citenamefont {Sethi}}]{Chang:2018dge}%
  \BibitemOpen
  \bibfield  {author} {\bibinfo {author} {\bibfnamefont {Chih-Kai}\
  \bibnamefont {Chang}}, \bibinfo {author} {\bibfnamefont {Christian}\
  \bibnamefont {Ferko}}, \ and\ \bibinfo {author} {\bibfnamefont {Savdeep}\
  \bibnamefont {Sethi}},\ }\bibfield  {title} {\enquote {\bibinfo {title}
  {{Supersymmetry and $ T\overline{T} $ deformations}},}\ }\href {\doibase
  10.1007/JHEP04(2019)131} {\bibfield  {journal} {\bibinfo  {journal} {JHEP}\
  }\textbf {\bibinfo {volume} {04}},\ \bibinfo {pages} {131} (\bibinfo {year}
  {2019})},\ \Eprint {http://arxiv.org/abs/1811.01895} {arXiv:1811.01895
  [hep-th]} \BibitemShut {NoStop}%
\bibitem [{\citenamefont {Chang}\ \emph {et~al.}(2020)\citenamefont {Chang},
  \citenamefont {Ferko}, \citenamefont {Sethi}, \citenamefont {Sfondrini},\
  and\ \citenamefont {Tartaglino-Mazzucchelli}}]{Chang:2019kiu}%
  \BibitemOpen
  \bibfield  {author} {\bibinfo {author} {\bibfnamefont {Chih-Kai}\
  \bibnamefont {Chang}}, \bibinfo {author} {\bibfnamefont {Christian}\
  \bibnamefont {Ferko}}, \bibinfo {author} {\bibfnamefont {Savdeep}\
  \bibnamefont {Sethi}}, \bibinfo {author} {\bibfnamefont {Alessandro}\
  \bibnamefont {Sfondrini}}, \ and\ \bibinfo {author} {\bibfnamefont
  {Gabriele}\ \bibnamefont {Tartaglino-Mazzucchelli}},\ }\bibfield  {title}
  {\enquote {\bibinfo {title} {{$T\bar{T}$ flows and (2,2) supersymmetry}},}\
  }\href {\doibase 10.1103/PhysRevD.101.026008} {\bibfield  {journal} {\bibinfo
   {journal} {Phys. Rev. D}\ }\textbf {\bibinfo {volume} {101}},\ \bibinfo
  {pages} {026008} (\bibinfo {year} {2020})},\ \Eprint
  {http://arxiv.org/abs/1906.00467} {arXiv:1906.00467 [hep-th]} \BibitemShut
  {NoStop}%
\bibitem [{\citenamefont {He}\ \emph {et~al.}(2020)\citenamefont {He},
  \citenamefont {Sun},\ and\ \citenamefont {Sun}}]{He:2019ahx}%
  \BibitemOpen
  \bibfield  {author} {\bibinfo {author} {\bibfnamefont {Song}\ \bibnamefont
  {He}}, \bibinfo {author} {\bibfnamefont {Jia-Rui}\ \bibnamefont {Sun}}, \
  and\ \bibinfo {author} {\bibfnamefont {Yuan}\ \bibnamefont {Sun}},\
  }\bibfield  {title} {\enquote {\bibinfo {title} {{The correlation function of
  (1,1) and (2,2) supersymmetric theories with $T\bar{T}$ deformation}},}\
  }\href {\doibase 10.1007/JHEP04(2020)100} {\bibfield  {journal} {\bibinfo
  {journal} {JHEP}\ }\textbf {\bibinfo {volume} {04}},\ \bibinfo {pages} {100}
  (\bibinfo {year} {2020})},\ \Eprint {http://arxiv.org/abs/1912.11461}
  {arXiv:1912.11461 [hep-th]} \BibitemShut {NoStop}%
\bibitem [{\citenamefont {Jiang}\ \emph {et~al.}(2019)\citenamefont {Jiang},
  \citenamefont {Sfondrini},\ and\ \citenamefont
  {Tartaglino-Mazzucchelli}}]{Jiang:2019hux}%
  \BibitemOpen
  \bibfield  {author} {\bibinfo {author} {\bibfnamefont {Hongliang}\
  \bibnamefont {Jiang}}, \bibinfo {author} {\bibfnamefont {Alessandro}\
  \bibnamefont {Sfondrini}}, \ and\ \bibinfo {author} {\bibfnamefont
  {Gabriele}\ \bibnamefont {Tartaglino-Mazzucchelli}},\ }\bibfield  {title}
  {\enquote {\bibinfo {title} {{$T\bar{T}$ deformations with
  $\mathcal{N}=(0,2)$ supersymmetry}},}\ }\href {\doibase
  10.1103/PhysRevD.100.046017} {\bibfield  {journal} {\bibinfo  {journal}
  {Phys. Rev. D}\ }\textbf {\bibinfo {volume} {100}},\ \bibinfo {pages}
  {046017} (\bibinfo {year} {2019})},\ \Eprint
  {http://arxiv.org/abs/1904.04760} {arXiv:1904.04760 [hep-th]} \BibitemShut
  {NoStop}%
\bibitem [{\citenamefont {Ferko}\ \emph {et~al.}(2020)\citenamefont {Ferko},
  \citenamefont {Jiang}, \citenamefont {Sethi},\ and\ \citenamefont
  {Tartaglino-Mazzucchelli}}]{Ferko:2019oyv}%
  \BibitemOpen
  \bibfield  {author} {\bibinfo {author} {\bibfnamefont {Christian}\
  \bibnamefont {Ferko}}, \bibinfo {author} {\bibfnamefont {Hongliang}\
  \bibnamefont {Jiang}}, \bibinfo {author} {\bibfnamefont {Savdeep}\
  \bibnamefont {Sethi}}, \ and\ \bibinfo {author} {\bibfnamefont {Gabriele}\
  \bibnamefont {Tartaglino-Mazzucchelli}},\ }\bibfield  {title} {\enquote
  {\bibinfo {title} {{Non-linear supersymmetry and $ T\overline{T} $-like
  flows}},}\ }\href {\doibase 10.1007/JHEP02(2020)016} {\bibfield  {journal}
  {\bibinfo  {journal} {JHEP}\ }\textbf {\bibinfo {volume} {02}},\ \bibinfo
  {pages} {016} (\bibinfo {year} {2020})},\ \Eprint
  {http://arxiv.org/abs/1910.01599} {arXiv:1910.01599 [hep-th]} \BibitemShut
  {NoStop}%
\bibitem [{\citenamefont {Cribiori}\ \emph {et~al.}(2019)\citenamefont
  {Cribiori}, \citenamefont {Farakos},\ and\ \citenamefont {von
  Unge}}]{Cribiori:2019xzp}%
  \BibitemOpen
  \bibfield  {author} {\bibinfo {author} {\bibfnamefont {Niccol\`o}\
  \bibnamefont {Cribiori}}, \bibinfo {author} {\bibfnamefont {Fotis}\
  \bibnamefont {Farakos}}, \ and\ \bibinfo {author} {\bibfnamefont {Rikard}\
  \bibnamefont {von Unge}},\ }\bibfield  {title} {\enquote {\bibinfo {title}
  {{2D Volkov-Akulov Model as a $T \overline{T}$ Deformation}},}\ }\href
  {\doibase 10.1103/PhysRevLett.123.201601} {\bibfield  {journal} {\bibinfo
  {journal} {Phys. Rev. Lett.}\ }\textbf {\bibinfo {volume} {123}},\ \bibinfo
  {pages} {201601} (\bibinfo {year} {2019})},\ \Eprint
  {http://arxiv.org/abs/1907.08150} {arXiv:1907.08150 [hep-th]} \BibitemShut
  {NoStop}%
\bibitem [{\citenamefont {Frolov}(2020{\natexlab{a}})}]{Frolov:2019nrr}%
  \BibitemOpen
  \bibfield  {author} {\bibinfo {author} {\bibfnamefont {Sergey}\ \bibnamefont
  {Frolov}},\ }\bibfield  {title} {\enquote {\bibinfo {title} {{$T\overline{T}$
  Deformation and the Light-Cone Gauge}},}\ }\href {\doibase
  10.1134/S0081543820030098} {\bibfield  {journal} {\bibinfo  {journal} {Proc.
  Steklov Inst. Math.}\ }\textbf {\bibinfo {volume} {309}},\ \bibinfo {pages}
  {107--126} (\bibinfo {year} {2020}{\natexlab{a}})},\ \Eprint
  {http://arxiv.org/abs/1905.07946} {arXiv:1905.07946 [hep-th]} \BibitemShut
  {NoStop}%
\bibitem [{\citenamefont {Frolov}(2020{\natexlab{b}})}]{Frolov:2019xzi}%
  \BibitemOpen
  \bibfield  {author} {\bibinfo {author} {\bibfnamefont {Sergey}\ \bibnamefont
  {Frolov}},\ }\bibfield  {title} {\enquote {\bibinfo {title} {{$T{\overline
  T}$, $\widetilde JJ$, $JT$ and $\widetilde JT$ deformations}},}\ }\href
  {\doibase 10.1088/1751-8121/ab581b} {\bibfield  {journal} {\bibinfo
  {journal} {J. Phys. A}\ }\textbf {\bibinfo {volume} {53}},\ \bibinfo {pages}
  {025401} (\bibinfo {year} {2020}{\natexlab{b}})},\ \Eprint
  {http://arxiv.org/abs/1907.12117} {arXiv:1907.12117 [hep-th]} \BibitemShut
  {NoStop}%
\bibitem [{\citenamefont {Mezincescu}\ and\ \citenamefont
  {Townsend}(2011)}]{Mezincescu:2011nh}%
  \BibitemOpen
  \bibfield  {author} {\bibinfo {author} {\bibfnamefont {Luca}\ \bibnamefont
  {Mezincescu}}\ and\ \bibinfo {author} {\bibfnamefont {Paul~K.}\ \bibnamefont
  {Townsend}},\ }\bibfield  {title} {\enquote {\bibinfo {title} {{Quantum 3D
  Superstrings}},}\ }\href {\doibase 10.1103/PhysRevD.84.106006} {\bibfield
  {journal} {\bibinfo  {journal} {Phys. Rev. D}\ }\textbf {\bibinfo {volume}
  {84}},\ \bibinfo {pages} {106006} (\bibinfo {year} {2011})},\ \Eprint
  {http://arxiv.org/abs/1106.1374} {arXiv:1106.1374 [hep-th]} \BibitemShut
  {NoStop}%
\bibitem [{\citenamefont {Ivanov}\ \emph {et~al.}(2001)\citenamefont {Ivanov},
  \citenamefont {Krivonos}, \citenamefont {Lechtenfeld},\ and\ \citenamefont
  {Zupnik}}]{Ivanov:2000nk}%
  \BibitemOpen
  \bibfield  {author} {\bibinfo {author} {\bibfnamefont {E.}~\bibnamefont
  {Ivanov}}, \bibinfo {author} {\bibfnamefont {S.}~\bibnamefont {Krivonos}},
  \bibinfo {author} {\bibfnamefont {O.}~\bibnamefont {Lechtenfeld}}, \ and\
  \bibinfo {author} {\bibfnamefont {B.}~\bibnamefont {Zupnik}},\ }\bibfield
  {title} {\enquote {\bibinfo {title} {{Partial spontaneous breaking of
  two-dimensional supersymmetry}},}\ }\href@noop {} {\bibfield  {journal}
  {\bibinfo  {journal} {Nucl. Phys. B}\ }\textbf {\bibinfo {volume} {600}},\
  \bibinfo {pages} {235--271} (\bibinfo {year} {2001})}\BibitemShut {NoStop}%
\bibitem [{\citenamefont {Umezawa}\ and\ \citenamefont
  {Kamefuchi}(1961)}]{Umezawa:1961igm}%
  \BibitemOpen
  \bibfield  {author} {\bibinfo {author} {\bibfnamefont {H.}~\bibnamefont
  {Umezawa}}\ and\ \bibinfo {author} {\bibfnamefont {S.}~\bibnamefont
  {Kamefuchi}},\ }\bibfield  {title} {\enquote {\bibinfo {title} {{Equivalence
  theorems and renormalization problem in vector field theory (The Yang-Mills
  field with non-vanishing masses)}},}\ }\href {\doibase
  10.1016/0029-5582(61)90269-3} {\bibfield  {journal} {\bibinfo  {journal}
  {Nucl. Phys.}\ }\textbf {\bibinfo {volume} {23}},\ \bibinfo {pages}
  {399--429} (\bibinfo {year} {1961})}\BibitemShut {NoStop}%
\bibitem [{\citenamefont {Kamefuchi}\ \emph {et~al.}(1961)\citenamefont
  {Kamefuchi}, \citenamefont {O'Raifeartaigh},\ and\ \citenamefont
  {Salam}}]{Kamefuchi:1961sb}%
  \BibitemOpen
  \bibfield  {author} {\bibinfo {author} {\bibfnamefont {S.}~\bibnamefont
  {Kamefuchi}}, \bibinfo {author} {\bibfnamefont {L.}~\bibnamefont
  {O'Raifeartaigh}}, \ and\ \bibinfo {author} {\bibfnamefont {Abdus}\
  \bibnamefont {Salam}},\ }\bibfield  {title} {\enquote {\bibinfo {title}
  {{Change of variables and equivalence theorems in quantum field theories}},}\
  }\href {\doibase 10.1016/0029-5582(61)90056-6} {\bibfield  {journal}
  {\bibinfo  {journal} {Nucl. Phys.}\ }\textbf {\bibinfo {volume} {28}},\
  \bibinfo {pages} {529--549} (\bibinfo {year} {1961})}\BibitemShut {NoStop}%
\bibitem [{\citenamefont {Salam}\ and\ \citenamefont
  {Strathdee}(1970)}]{Salam:1970fso}%
  \BibitemOpen
  \bibfield  {author} {\bibinfo {author} {\bibfnamefont {Abdus}\ \bibnamefont
  {Salam}}\ and\ \bibinfo {author} {\bibfnamefont {J.~A.}\ \bibnamefont
  {Strathdee}},\ }\bibfield  {title} {\enquote {\bibinfo {title} {{Equivalent
  formulations of massive vector field theories}},}\ }\href {\doibase
  10.1103/PhysRevD.2.2869} {\bibfield  {journal} {\bibinfo  {journal} {Phys.
  Rev. D}\ }\textbf {\bibinfo {volume} {2}},\ \bibinfo {pages} {2869--2876}
  (\bibinfo {year} {1970})}\BibitemShut {NoStop}%
\bibitem [{\citenamefont {Slavnov}(1991)}]{Slavnov:1990nd}%
  \BibitemOpen
  \bibfield  {author} {\bibinfo {author} {\bibfnamefont {A.~A.}\ \bibnamefont
  {Slavnov}},\ }\bibfield  {title} {\enquote {\bibinfo {title} {{Equivalence
  theorem for spectrum changing transformations}},}\ }\href {\doibase
  10.1016/0370-2693(91)91105-5} {\bibfield  {journal} {\bibinfo  {journal}
  {Phys. Lett. B}\ }\textbf {\bibinfo {volume} {258}},\ \bibinfo {pages}
  {391--394} (\bibinfo {year} {1991})}\BibitemShut {NoStop}%
\bibitem [{\citenamefont {Bastianelli}(1991)}]{Bastianelli:1990ey}%
  \BibitemOpen
  \bibfield  {author} {\bibinfo {author} {\bibfnamefont {Fiorenzo}\
  \bibnamefont {Bastianelli}},\ }\bibfield  {title} {\enquote {\bibinfo {title}
  {{BRST symmetry from a change of variables and the gauged WZNW models}},}\
  }\href {\doibase 10.1016/0550-3213(91)90273-Z} {\bibfield  {journal}
  {\bibinfo  {journal} {Nucl. Phys. B}\ }\textbf {\bibinfo {volume} {361}},\
  \bibinfo {pages} {555--567} (\bibinfo {year} {1991})}\BibitemShut {NoStop}%
\bibitem [{\citenamefont {Alfaro}\ and\ \citenamefont
  {Damgaard}(1992)}]{Alfaro:1992np}%
  \BibitemOpen
  \bibfield  {author} {\bibinfo {author} {\bibfnamefont {J.}~\bibnamefont
  {Alfaro}}\ and\ \bibinfo {author} {\bibfnamefont {P.~H.}\ \bibnamefont
  {Damgaard}},\ }\bibfield  {title} {\enquote {\bibinfo {title} {{BRST symmetry
  of field redefinitions}},}\ }\href {\doibase 10.1016/0003-4916(92)90360-X}
  {\bibfield  {journal} {\bibinfo  {journal} {Annals Phys.}\ }\textbf {\bibinfo
  {volume} {220}},\ \bibinfo {pages} {188--211} (\bibinfo {year}
  {1992})}\BibitemShut {NoStop}%
\bibitem [{\citenamefont {Kausch}(1995)}]{Kausch:1995py}%
  \BibitemOpen
  \bibfield  {author} {\bibinfo {author} {\bibfnamefont {Horst~G.}\
  \bibnamefont {Kausch}},\ }\bibfield  {title} {\enquote {\bibinfo {title}
  {{Curiosities at c = -2}},}\ }\href@noop {} {\  (\bibinfo {year} {1995})},\
  \Eprint {http://arxiv.org/abs/hep-th/9510149} {arXiv:hep-th/9510149}
  \BibitemShut {NoStop}%
\bibitem [{\citenamefont {Gaberdiel}\ and\ \citenamefont
  {Kausch}(1999)}]{Gaberdiel:1998ps}%
  \BibitemOpen
  \bibfield  {author} {\bibinfo {author} {\bibfnamefont {Matthias~R.}\
  \bibnamefont {Gaberdiel}}\ and\ \bibinfo {author} {\bibfnamefont {Horst~G.}\
  \bibnamefont {Kausch}},\ }\bibfield  {title} {\enquote {\bibinfo {title} {{A
  Local logarithmic conformal field theory}},}\ }\href {\doibase
  10.1016/S0550-3213(98)00701-9} {\bibfield  {journal} {\bibinfo  {journal}
  {Nucl. Phys. B}\ }\textbf {\bibinfo {volume} {538}},\ \bibinfo {pages}
  {631--658} (\bibinfo {year} {1999})},\ \Eprint
  {http://arxiv.org/abs/hep-th/9807091} {arXiv:hep-th/9807091} \BibitemShut
  {NoStop}%
\bibitem [{\citenamefont {Kausch}(2000)}]{Kausch:2000fu}%
  \BibitemOpen
  \bibfield  {author} {\bibinfo {author} {\bibfnamefont {Horst~G.}\
  \bibnamefont {Kausch}},\ }\bibfield  {title} {\enquote {\bibinfo {title}
  {{Symplectic fermions}},}\ }\href {\doibase 10.1016/S0550-3213(00)00295-9}
  {\bibfield  {journal} {\bibinfo  {journal} {Nucl. Phys. B}\ }\textbf
  {\bibinfo {volume} {583}},\ \bibinfo {pages} {513--541} (\bibinfo {year}
  {2000})},\ \Eprint {http://arxiv.org/abs/hep-th/0003029}
  {arXiv:hep-th/0003029} \BibitemShut {NoStop}%
\bibitem [{\citenamefont {LeClair}\ and\ \citenamefont
  {Neubert}(2007)}]{LeClair:2007iy}%
  \BibitemOpen
  \bibfield  {author} {\bibinfo {author} {\bibfnamefont {Andre}\ \bibnamefont
  {LeClair}}\ and\ \bibinfo {author} {\bibfnamefont {Matthias}\ \bibnamefont
  {Neubert}},\ }\bibfield  {title} {\enquote {\bibinfo {title} {{Semi-Lorentz
  invariance, unitarity, and critical exponents of symplectic fermion
  models}},}\ }\href {\doibase 10.1088/1126-6708/2007/10/027} {\bibfield
  {journal} {\bibinfo  {journal} {JHEP}\ }\textbf {\bibinfo {volume} {10}},\
  \bibinfo {pages} {027} (\bibinfo {year} {2007})},\ \Eprint
  {http://arxiv.org/abs/0705.4657} {arXiv:0705.4657 [hep-th]} \BibitemShut
  {NoStop}%
\bibitem [{\citenamefont {Robinson}\ \emph {et~al.}(2009)\citenamefont
  {Robinson}, \citenamefont {Kapit},\ and\ \citenamefont
  {LeClair}}]{Robinson:2009xm}%
  \BibitemOpen
  \bibfield  {author} {\bibinfo {author} {\bibfnamefont {Dean~J.}\ \bibnamefont
  {Robinson}}, \bibinfo {author} {\bibfnamefont {Eliot}\ \bibnamefont {Kapit}},
  \ and\ \bibinfo {author} {\bibfnamefont {Andre}\ \bibnamefont {LeClair}},\
  }\bibfield  {title} {\enquote {\bibinfo {title} {{Lorentz Symmetric Quantum
  Field Theory for Symplectic Fermions}},}\ }\href {\doibase 10.1063/1.3248256}
  {\bibfield  {journal} {\bibinfo  {journal} {J. Math. Phys.}\ }\textbf
  {\bibinfo {volume} {50}},\ \bibinfo {pages} {112301} (\bibinfo {year}
  {2009})},\ \Eprint {http://arxiv.org/abs/0903.2399} {arXiv:0903.2399
  [hep-th]} \BibitemShut {NoStop}%
\bibitem [{\citenamefont {Ryu}\ and\ \citenamefont {Yoon}(2022)}]{Ryu:2022ffg}%
  \BibitemOpen
  \bibfield  {author} {\bibinfo {author} {\bibfnamefont {Shinsei}\ \bibnamefont
  {Ryu}}\ and\ \bibinfo {author} {\bibfnamefont {Junggi}\ \bibnamefont
  {Yoon}},\ }\bibfield  {title} {\enquote {\bibinfo {title} {{Unitarity of
  Symplectic Fermion in $\alpha$-vacua with Negative Central Charge}},}\
  }\href@noop {} {\  (\bibinfo {year} {2022})},\ \Eprint
  {http://arxiv.org/abs/2208.12169} {arXiv:2208.12169 [hep-th]} \BibitemShut
  {NoStop}%
\bibitem [{\citenamefont {Conti}\ \emph {et~al.}(2018)\citenamefont {Conti},
  \citenamefont {Iannella}, \citenamefont {Negro},\ and\ \citenamefont
  {Tateo}}]{Conti:2018jho}%
  \BibitemOpen
  \bibfield  {author} {\bibinfo {author} {\bibfnamefont {Riccardo}\
  \bibnamefont {Conti}}, \bibinfo {author} {\bibfnamefont {Leonardo}\
  \bibnamefont {Iannella}}, \bibinfo {author} {\bibfnamefont {Stefano}\
  \bibnamefont {Negro}}, \ and\ \bibinfo {author} {\bibfnamefont {Roberto}\
  \bibnamefont {Tateo}},\ }\bibfield  {title} {\enquote {\bibinfo {title}
  {{Generalised Born-Infeld models, Lax operators and the $
  \mathrm{T}\overline{\mathrm{T}} $ perturbation}},}\ }\href {\doibase
  10.1007/JHEP11(2018)007} {\bibfield  {journal} {\bibinfo  {journal} {JHEP}\
  }\textbf {\bibinfo {volume} {11}},\ \bibinfo {pages} {007} (\bibinfo {year}
  {2018})},\ \Eprint {http://arxiv.org/abs/1806.11515} {arXiv:1806.11515
  [hep-th]} \BibitemShut {NoStop}%
\bibitem [{\citenamefont {Nielsen}\ and\ \citenamefont
  {Olesen}(1973)}]{Nielsen:1973cs}%
  \BibitemOpen
  \bibfield  {author} {\bibinfo {author} {\bibfnamefont {Holger~Bech}\
  \bibnamefont {Nielsen}}\ and\ \bibinfo {author} {\bibfnamefont
  {P.}~\bibnamefont {Olesen}},\ }\bibfield  {title} {\enquote {\bibinfo {title}
  {{Vortex Line Models for Dual Strings}},}\ }\href {\doibase
  10.1016/0550-3213(73)90350-7} {\bibfield  {journal} {\bibinfo  {journal}
  {Nucl. Phys. B}\ }\textbf {\bibinfo {volume} {61}},\ \bibinfo {pages}
  {45--61} (\bibinfo {year} {1973})}\BibitemShut {NoStop}%
\bibitem [{\citenamefont {Volkov}\ and\ \citenamefont
  {Akulov}(1973)}]{Volkov:1973ix}%
  \BibitemOpen
  \bibfield  {author} {\bibinfo {author} {\bibfnamefont {D.~V.}\ \bibnamefont
  {Volkov}}\ and\ \bibinfo {author} {\bibfnamefont {V.~P.}\ \bibnamefont
  {Akulov}},\ }\bibfield  {title} {\enquote {\bibinfo {title} {{Is the Neutrino
  a Goldstone Particle?}}}\ }\href {\doibase 10.1016/0370-2693(73)90490-5}
  {\bibfield  {journal} {\bibinfo  {journal} {Phys. Lett. B}\ }\textbf
  {\bibinfo {volume} {46}},\ \bibinfo {pages} {109--110} (\bibinfo {year}
  {1973})}\BibitemShut {NoStop}%
\bibitem [{\citenamefont {Volkov}(1973)}]{Volkov:1973vd}%
  \BibitemOpen
  \bibfield  {author} {\bibinfo {author} {\bibfnamefont {Dmitri~V.}\
  \bibnamefont {Volkov}},\ }\bibfield  {title} {\enquote {\bibinfo {title}
  {{Phenomenological Lagrangians}},}\ }\href@noop {} {\bibfield  {journal}
  {\bibinfo  {journal} {Fiz. Elem. Chast. Atom. Yadra}\ }\textbf {\bibinfo
  {volume} {4}},\ \bibinfo {pages} {3--41} (\bibinfo {year}
  {1973})}\BibitemShut {NoStop}%
\bibitem [{\citenamefont {Ogievetsky}(1974)}]{Ogievetsky:1974}%
  \BibitemOpen
  \bibfield  {author} {\bibinfo {author} {\bibfnamefont {Viktor~I.}\
  \bibnamefont {Ogievetsky}},\ }\bibfield  {title} {\enquote {\bibinfo {title}
  {{Nonlinear realizations of internal and space-time symmetries.}}}\
  }\href@noop {} {\  (\bibinfo {year} {1974})}\BibitemShut {NoStop}%
\bibitem [{\citenamefont {Penco}(2020)}]{Penco:2020kvy}%
  \BibitemOpen
  \bibfield  {author} {\bibinfo {author} {\bibfnamefont {Riccardo}\
  \bibnamefont {Penco}},\ }\bibfield  {title} {\enquote {\bibinfo {title} {{An
  Introduction to Effective Field Theories}},}\ }\href@noop {} {\  (\bibinfo
  {year} {2020})},\ \Eprint {http://arxiv.org/abs/2006.16285} {arXiv:2006.16285
  [hep-th]} \BibitemShut {NoStop}%
\bibitem [{\citenamefont {Rocek}(1978)}]{Rocek:1978nb}%
  \BibitemOpen
  \bibfield  {author} {\bibinfo {author} {\bibfnamefont {M.}~\bibnamefont
  {Rocek}},\ }\bibfield  {title} {\enquote {\bibinfo {title} {{Linearizing the
  Volkov-Akulov Model}},}\ }\href {\doibase 10.1103/PhysRevLett.41.451}
  {\bibfield  {journal} {\bibinfo  {journal} {Phys. Rev. Lett.}\ }\textbf
  {\bibinfo {volume} {41}},\ \bibinfo {pages} {451--453} (\bibinfo {year}
  {1978})}\BibitemShut {NoStop}%
\bibitem [{\citenamefont {Komargodski}\ and\ \citenamefont
  {Seiberg}(2009)}]{Komargodski:2009rz}%
  \BibitemOpen
  \bibfield  {author} {\bibinfo {author} {\bibfnamefont {Zohar}\ \bibnamefont
  {Komargodski}}\ and\ \bibinfo {author} {\bibfnamefont {Nathan}\ \bibnamefont
  {Seiberg}},\ }\bibfield  {title} {\enquote {\bibinfo {title} {{From Linear
  SUSY to Constrained Superfields}},}\ }\href {\doibase
  10.1088/1126-6708/2009/09/066} {\bibfield  {journal} {\bibinfo  {journal}
  {JHEP}\ }\textbf {\bibinfo {volume} {09}},\ \bibinfo {pages} {066} (\bibinfo
  {year} {2009})},\ \Eprint {http://arxiv.org/abs/0907.2441} {arXiv:0907.2441
  [hep-th]} \BibitemShut {NoStop}%
\bibitem [{\citenamefont {Bagger}\ and\ \citenamefont
  {Galperin}(1997)}]{Bagger:1996wp}%
  \BibitemOpen
  \bibfield  {author} {\bibinfo {author} {\bibfnamefont {Jonathan}\
  \bibnamefont {Bagger}}\ and\ \bibinfo {author} {\bibfnamefont {Alexander}\
  \bibnamefont {Galperin}},\ }\bibfield  {title} {\enquote {\bibinfo {title}
  {{A New Goldstone multiplet for partially broken supersymmetry}},}\ }\href
  {\doibase 10.1103/PhysRevD.55.1091} {\bibfield  {journal} {\bibinfo
  {journal} {Phys. Rev. D}\ }\textbf {\bibinfo {volume} {55}},\ \bibinfo
  {pages} {1091--1098} (\bibinfo {year} {1997})},\ \Eprint
  {http://arxiv.org/abs/hep-th/9608177} {arXiv:hep-th/9608177} \BibitemShut
  {NoStop}%
\bibitem [{\citenamefont {Coleman}\ \emph {et~al.}(1969)\citenamefont
  {Coleman}, \citenamefont {Wess},\ and\ \citenamefont
  {Zumino}}]{Coleman:1969sm}%
  \BibitemOpen
  \bibfield  {author} {\bibinfo {author} {\bibfnamefont {Sidney~R.}\
  \bibnamefont {Coleman}}, \bibinfo {author} {\bibfnamefont {J.}~\bibnamefont
  {Wess}}, \ and\ \bibinfo {author} {\bibfnamefont {Bruno}\ \bibnamefont
  {Zumino}},\ }\bibfield  {title} {\enquote {\bibinfo {title} {{Structure of
  phenomenological Lagrangians. 1.}}}\ }\href {\doibase
  10.1103/PhysRev.177.2239} {\bibfield  {journal} {\bibinfo  {journal} {Phys.
  Rev.}\ }\textbf {\bibinfo {volume} {177}},\ \bibinfo {pages} {2239--2247}
  (\bibinfo {year} {1969})}\BibitemShut {NoStop}%
\bibitem [{\citenamefont {Callan}\ \emph {et~al.}(1969)\citenamefont {Callan},
  \citenamefont {Coleman}, \citenamefont {Wess},\ and\ \citenamefont
  {Zumino}}]{Callan:1969sn}%
  \BibitemOpen
  \bibfield  {author} {\bibinfo {author} {\bibfnamefont {Curtis~G.}\
  \bibnamefont {Callan}, \bibfnamefont {Jr.}}, \bibinfo {author} {\bibfnamefont
  {Sidney~R.}\ \bibnamefont {Coleman}}, \bibinfo {author} {\bibfnamefont
  {J.}~\bibnamefont {Wess}}, \ and\ \bibinfo {author} {\bibfnamefont {Bruno}\
  \bibnamefont {Zumino}},\ }\bibfield  {title} {\enquote {\bibinfo {title}
  {{Structure of phenomenological Lagrangians. 2.}}}\ }\href {\doibase
  10.1103/PhysRev.177.2247} {\bibfield  {journal} {\bibinfo  {journal} {Phys.
  Rev.}\ }\textbf {\bibinfo {volume} {177}},\ \bibinfo {pages} {2247--2250}
  (\bibinfo {year} {1969})}\BibitemShut {NoStop}%
\bibitem [{\citenamefont {Ivanov}\ and\ \citenamefont
  {Ogievetsky}(1975)}]{Ivanov:1975zq}%
  \BibitemOpen
  \bibfield  {author} {\bibinfo {author} {\bibfnamefont {E.~A.}\ \bibnamefont
  {Ivanov}}\ and\ \bibinfo {author} {\bibfnamefont {V.~I.}\ \bibnamefont
  {Ogievetsky}},\ }\bibfield  {title} {\enquote {\bibinfo {title} {{The Inverse
  Higgs Phenomenon in Nonlinear Realizations}},}\ }\href {\doibase
  10.1007/BF01028947} {\bibfield  {journal} {\bibinfo  {journal} {Teor. Mat.
  Fiz.}\ }\textbf {\bibinfo {volume} {25}},\ \bibinfo {pages} {164--177}
  (\bibinfo {year} {1975})}\BibitemShut {NoStop}%
\bibitem [{\citenamefont {Gomis}\ \emph {et~al.}(2006)\citenamefont {Gomis},
  \citenamefont {Kamimura},\ and\ \citenamefont {West}}]{Gomis:2006wu}%
  \BibitemOpen
  \bibfield  {author} {\bibinfo {author} {\bibfnamefont {Joaquim}\ \bibnamefont
  {Gomis}}, \bibinfo {author} {\bibfnamefont {Kiyoshi}\ \bibnamefont
  {Kamimura}}, \ and\ \bibinfo {author} {\bibfnamefont {Peter~C.}\ \bibnamefont
  {West}},\ }\bibfield  {title} {\enquote {\bibinfo {title} {{Diffeomorphism,
  kappa transformations and the theory of non-linear realisations}},}\ }\href
  {\doibase 10.1088/1126-6708/2006/10/015} {\bibfield  {journal} {\bibinfo
  {journal} {JHEP}\ }\textbf {\bibinfo {volume} {10}},\ \bibinfo {pages} {015}
  (\bibinfo {year} {2006})},\ \Eprint {http://arxiv.org/abs/hep-th/0607104}
  {arXiv:hep-th/0607104} \BibitemShut {NoStop}%
\bibitem [{\citenamefont {Zheltukhin}(2010)}]{Zheltukhin:2010xr}%
  \BibitemOpen
  \bibfield  {author} {\bibinfo {author} {\bibfnamefont {A.~A.}\ \bibnamefont
  {Zheltukhin}},\ }\bibfield  {title} {\enquote {\bibinfo {title} {{On
  Equivalence of the Komargodski-Seiberg Action to the Volkov-Akulov
  Action}},}\ }\href@noop {} {\  (\bibinfo {year} {2010})},\ \Eprint
  {http://arxiv.org/abs/1009.2166} {arXiv:1009.2166 [hep-th]} \BibitemShut
  {NoStop}%
\bibitem [{\citenamefont {Kuzenko}\ and\ \citenamefont
  {Tyler}(2011{\natexlab{a}})}]{Kuzenko:2010ef}%
  \BibitemOpen
  \bibfield  {author} {\bibinfo {author} {\bibfnamefont {Sergei~M.}\
  \bibnamefont {Kuzenko}}\ and\ \bibinfo {author} {\bibfnamefont {Simon~J.}\
  \bibnamefont {Tyler}},\ }\bibfield  {title} {\enquote {\bibinfo {title}
  {{Relating the Komargodski-Seiberg and Akulov-Volkov actions: Exact nonlinear
  field redefinition}},}\ }\href {\doibase 10.1016/j.physletb.2011.03.020}
  {\bibfield  {journal} {\bibinfo  {journal} {Phys. Lett. B}\ }\textbf
  {\bibinfo {volume} {698}},\ \bibinfo {pages} {319--322} (\bibinfo {year}
  {2011}{\natexlab{a}})},\ \Eprint {http://arxiv.org/abs/1009.3298}
  {arXiv:1009.3298 [hep-th]} \BibitemShut {NoStop}%
\bibitem [{\citenamefont {Kuzenko}\ and\ \citenamefont
  {Tyler}(2011{\natexlab{b}})}]{Kuzenko:2011tj}%
  \BibitemOpen
  \bibfield  {author} {\bibinfo {author} {\bibfnamefont {Sergei~M.}\
  \bibnamefont {Kuzenko}}\ and\ \bibinfo {author} {\bibfnamefont {Simon~J.}\
  \bibnamefont {Tyler}},\ }\bibfield  {title} {\enquote {\bibinfo {title} {{On
  the Goldstino actions and their symmetries}},}\ }\href {\doibase
  10.1007/JHEP05(2011)055} {\bibfield  {journal} {\bibinfo  {journal} {JHEP}\
  }\textbf {\bibinfo {volume} {05}},\ \bibinfo {pages} {055} (\bibinfo {year}
  {2011}{\natexlab{b}})},\ \Eprint {http://arxiv.org/abs/1102.3043}
  {arXiv:1102.3043 [hep-th]} \BibitemShut {NoStop}%
\bibitem [{\citenamefont {Dubovsky}\ \emph {et~al.}(2012)\citenamefont
  {Dubovsky}, \citenamefont {Flauger},\ and\ \citenamefont
  {Gorbenko}}]{Dubovsky:2012sh}%
  \BibitemOpen
  \bibfield  {author} {\bibinfo {author} {\bibfnamefont {Sergei}\ \bibnamefont
  {Dubovsky}}, \bibinfo {author} {\bibfnamefont {Raphael}\ \bibnamefont
  {Flauger}}, \ and\ \bibinfo {author} {\bibfnamefont {Victor}\ \bibnamefont
  {Gorbenko}},\ }\bibfield  {title} {\enquote {\bibinfo {title} {{Effective
  String Theory Revisited}},}\ }\href {\doibase 10.1007/JHEP09(2012)044}
  {\bibfield  {journal} {\bibinfo  {journal} {JHEP}\ }\textbf {\bibinfo
  {volume} {09}},\ \bibinfo {pages} {044} (\bibinfo {year} {2012})},\ \Eprint
  {http://arxiv.org/abs/1203.1054} {arXiv:1203.1054 [hep-th]} \BibitemShut
  {NoStop}%
\bibitem [{\citenamefont {Dubovsky}\ \emph {et~al.}(2017)\citenamefont
  {Dubovsky}, \citenamefont {Gorbenko},\ and\ \citenamefont
  {Mirbabayi}}]{Dubovsky:2017cnj}%
  \BibitemOpen
  \bibfield  {author} {\bibinfo {author} {\bibfnamefont {Sergei}\ \bibnamefont
  {Dubovsky}}, \bibinfo {author} {\bibfnamefont {Victor}\ \bibnamefont
  {Gorbenko}}, \ and\ \bibinfo {author} {\bibfnamefont {Mehrdad}\ \bibnamefont
  {Mirbabayi}},\ }\bibfield  {title} {\enquote {\bibinfo {title} {{Asymptotic
  fragility, near AdS$_{2}$ holography and $ T\overline{T} $}},}\ }\href
  {\doibase 10.1007/JHEP09(2017)136} {\bibfield  {journal} {\bibinfo  {journal}
  {JHEP}\ }\textbf {\bibinfo {volume} {09}},\ \bibinfo {pages} {136} (\bibinfo
  {year} {2017})},\ \Eprint {http://arxiv.org/abs/1706.06604} {arXiv:1706.06604
  [hep-th]} \BibitemShut {NoStop}%
\bibitem [{\citenamefont {Dubovsky}\ \emph {et~al.}(2018)\citenamefont
  {Dubovsky}, \citenamefont {Gorbenko},\ and\ \citenamefont
  {Hern\'andez-Chifflet}}]{Dubovsky:2018bmo}%
  \BibitemOpen
  \bibfield  {author} {\bibinfo {author} {\bibfnamefont {Sergei}\ \bibnamefont
  {Dubovsky}}, \bibinfo {author} {\bibfnamefont {Victor}\ \bibnamefont
  {Gorbenko}}, \ and\ \bibinfo {author} {\bibfnamefont {Guzm\'an}\ \bibnamefont
  {Hern\'andez-Chifflet}},\ }\bibfield  {title} {\enquote {\bibinfo {title} {{$
  T\overline{T} $ partition function from topological gravity}},}\ }\href
  {\doibase 10.1007/JHEP09(2018)158} {\bibfield  {journal} {\bibinfo  {journal}
  {JHEP}\ }\textbf {\bibinfo {volume} {09}},\ \bibinfo {pages} {158} (\bibinfo
  {year} {2018})},\ \Eprint {http://arxiv.org/abs/1805.07386} {arXiv:1805.07386
  [hep-th]} \BibitemShut {NoStop}%
\bibitem [{\citenamefont {He}\ \emph {et~al.}(2023)\citenamefont {He},
  \citenamefont {Ouyang},\ and\ \citenamefont {Sun}}]{He:2022bbb}%
  \BibitemOpen
  \bibfield  {author} {\bibinfo {author} {\bibfnamefont {Song}\ \bibnamefont
  {He}}, \bibinfo {author} {\bibfnamefont {Hao}\ \bibnamefont {Ouyang}}, \ and\
  \bibinfo {author} {\bibfnamefont {Yuan}\ \bibnamefont {Sun}},\ }\bibfield
  {title} {\enquote {\bibinfo {title} {{Note on $T{\bar{T}}$ deformed matrix
  models and JT supergravity duals}},}\ }\href {\doibase
  10.1140/epjc/s10052-023-12019-3} {\bibfield  {journal} {\bibinfo  {journal}
  {Eur. Phys. J. C}\ }\textbf {\bibinfo {volume} {83}},\ \bibinfo {pages} {885}
  (\bibinfo {year} {2023})},\ \Eprint {http://arxiv.org/abs/2204.13636}
  {arXiv:2204.13636 [hep-th]} \BibitemShut {NoStop}%
\end{thebibliography}%

\end{document}